\documentclass[12pt]{article}
\usepackage[a4paper, left=0.75in, right=0.75in, top=1in, bottom=1in]{geometry}
\usepackage{graphicx}
\usepackage{amsmath}
\usepackage{amssymb}
\usepackage[authoryear,round]{natbib}
\usepackage{xcolor}
\usepackage{authblk}
\usepackage{subcaption}
\usepackage{lmodern}
\usepackage{setspace}
\usepackage{wrapfig}

\usepackage[hypertexnames=false, linktocpage=false, hidelinks]{hyperref} %

\title{A Spatiotemporal, Quasi-experimental Causal Inference Approach to Characterize the Effects of Global Plastic Waste Export and Burning on Air Quality Using Remotely Sensed Data} 
\author[*1,2]{Ellen M. Considine} 
\author[3]{Rachel C. Nethery}
\affil[1]{Cooperative Institute for Research in Environmental Sciences (CIRES), University of Colorado Boulder}
\affil[2]{Department of Geography, University of Colorado Boulder}
\affil[3]{Department of Biostatistics, Harvard T.H. Chan School of Public Health}
\affil[*]{Email: ellen.considine@colorado.edu}
\date{} %

\begin{document}

\maketitle

\abstract{Open burning of plastic waste may pose a significant threat to global health by degrading air quality, but quantitative research on this problem -- crucial for policy making -- has been stunted by lack of data. Many low- and middle-income countries, where open burning is most concerning, have little to no air quality monitoring. Here, we leverage remotely sensed data products combined with spatiotemporal causal analytic techniques to evaluate the impact of large-scale plastic waste policies on air quality. Throughout, we study Indonesia before and after 2018, when China halted its import of plastic waste, resulting in diversion of this massive waste stream to other countries. We tailor cutting-edge statistical methods to this setting, estimating effects of increased plastic waste imports on fine particulate matter (PM$_{2.5}$) near waste dump sites in Indonesia as a function of proximity to ports, an induced continuous exposure. We observe strong evidence that monthly PM$_{2.5}$increased after China's ban (2018-2019) relative to expected business-as-usual (2012-2017), with increases up to 1.68 $\mu$g/m$^3$ (95\% CI = [0.72, 2.48]) when exposed to medium-high port proximity. Effects were more modest for very high port proximity exposure, possibly reflecting smaller increases in dumping/burning where government oversight is greater.} %


\section{Introduction}\label{intro}

The open burning of waste, including plastic waste, is an urgent global health issue \citep{global_health_2024}. Open burning is defined as burning ``in open fires without managing for the emission of byproducts, such as gases and ash, into the ambient air or soil" \citep{pathak_2023_def}. An estimated 40-65\% of total municipal solid waste is open-burned in low- and middle-income countries (LMICs), largely as a result of two billion people around the world receiving no municipal solid waste collection \citep{global_health_2024}. As will be highlighted in this paper, LMICs also receive large amounts of waste from high-income countries.
Regardless of the source of the waste, waste burning threatens public health due to the air pollutants emitted, such as fine particulate matter  \citep{wiedinmyer_2014, guatemala_plastic_2023}, abbreviated PM$_{2.5}$. There is a large body of scientific evidence linking air pollution, and specifically PM$_{2.5}$, to health consequences ranging from respiratory and cardiovascular disease, to cancer, to reproductive and neurological disorders, to mortality \citep{niehs_air_pollution}. 

A decade ago, \citet{wiedinmyer_2014} presented the first comprehensive, global estimates of air pollution (and greenhouse gas) emissions due to open waste burning, using an emissions factor approach based on guidelines from the IPCC (Intergovernmental Panel on Climate Change). This approach relies on rough country-level approximations for the amount of waste generated and open-burned, which were derived as a deterministic function of national waste generation rate and waste collection efficiency estimates. 
The estimated amount of open waste burned was then converted into waste burning-attributable emissions estimates through multiplying by a literature-informed emissions factor. 
This and other inventories have subsequently been paired with chemical transport modeling to estimate the impact of open burning on regional air quality and attributable mortality \citep{africa_burning_emissions_2023}. 
While these studies provide evidence that the air pollution contributions of waste burning may be substantial, their methods require strong assumptions backed up by very limited empirical data (e.g., that all countries around the world burn approximately 60\% of the waste that is available to be burned), and adaptation of these methods to conduct policy impact evaluations, which are crucial for future policy making, would necessitate further strong assumptions about the impact of a policy on the various inputs.

In this study, we focus on plastic waste, whose global generation more than doubled from 2000-2019 \citep{oecd2022plastic} and whose transport / trade from higher- to lower-income countries is a major environmental justice issue. 
In contrast to earlier studies estimating emissions from burning all kinds of waste, recent studies have focused on more robust estimation of plastic waste generation in specific regions \citep{guatemala_plastic_2023} and detecting molecular tracers from plastic burning \citep{bangladesh_plastic_AQ_2022}. However, these methods do not easily scale to large regions and/or time periods. For context, \citet{guatemala_plastic_2023} surveyed 50 households in one rural indigenous community in Guatemala over 4 weeks and \citet{bangladesh_plastic_AQ_2022} deployed an air quality monitor sequentially at two sites in Bangladesh, leaving it in place for 3-4 months at each site.

To generate robust empirical evidence about the impacts of plastic waste burning on air quality at scale, here we leverage a case study featuring a large plastic waste quasi-experiment. On January 1, 2018, China banned the import of plastic waste\footnote{China also implemented a second ban in January 2019, tightening its initial policy. However, \citet{china_ban_AQ_2022} found that the impact of this second ban was minimal.}. Previously, between 1988 and 2016, they received 45\% of global plastic waste exports \citep{brooks_import_ban}. The 38 OECD (Organisation for Economic Co-operation and Development) countries, most of which are classified as high-income, contributed 64\% to all exports in this time frame, with the United States, Japan, and Germany topping the list \citep{brooks_import_ban}.
Unsurprisingly, in the wake of China's ban there was a marked increase in plastic waste exports to other countries in the East Asia \& Pacific region, which were already receiving 30\% of global exports \citep{brooks_import_ban}. 
In this paper, we focus on Indonesia, which in 2018 became a net importer of plastic waste \citep{NPAP_report}. As of 2020, it was estimated that 48\% of Indonesia's plastic waste is openly burned \citep{NPAP_report}. %

In this paper, we seek to quantify the increase in air pollution at open dump sites in Indonesia during 2018-2019 due to (the burning of) increased plastic waste in the wake of China's 2018 ban. However, data scarcity presents significant obstacles to such an analysis. Indonesia, like many LMICs, has very limited ground-level (``in situ") air quality monitoring. The country's first reliable air quality monitors were deployed in Jakarta during our study period \citep{jakarta_post_2016}.
For context: as of 2020, 51\% of the world's governments did not produce air pollution data \citep{openaq_report}\footnote{Note that OpenAQ has reported updates to this statistic in 2022 and 2024.}.
In recent years, remotely sensed data have been widely used to fill gaps in in-situ monitoring \citep{remote_sensing_aq}.
In addition to the lack of air quality data, to our knowledge there is no detailed record of Indonesia's plastic waste imports along the lines of what \citet{china_ban_AQ_2022} used to investigate domestic air quality impacts of China's ban -- and in any case, use of an official record might obscure the contribution of informal and/or covert transportation of waste, which is suggested by analysis of the Indonesian waste sector \citep{nicholas_institute_report}. Therefore, we develop an analytic strategy relying on data products derived from remote sensing, including PM$_{2.5}$ estimates, the identification of locations where waste has been openly dumped, and proximity to ports from which plastic waste can enter Indonesia (data sources are detailed in Section \ref{data}). To our knowledge, this is the first analysis providing empirical estimates of how a policy-driven change in plastic waste quantity -- and associated changes in burning -- impacts air quality, in a setting without in-situ air quality monitoring or waste quantity data.

Motivated by this case study, we also propose a new, multiply-robust causal inference approach to estimation and inference for quasi-experimental study designs with a ``universal'' but dose-dependent intervention. A growing body of work uses quasi-experimental study designs to estimate the air quality impact of a policy or intervention, with a recent surge focused on the Covid-19 lockdown -- which resembles our setting in causing a nationwide shift in emissions-relevant behaviors \citep{heffernan_2024,lee_chen_covid_2021,dey2021counterfactual}. Policies that are adopted ``universally'' across a country/region of interest present a challenge for many conventional quasi-experimental analytic methods, which typically assume the existence of a control group.
To address this, \citet{heffernan_2024} developed a framework for machine learning-based comparative interrupted time series (CITS), in which U.S. city-specific models were used to estimate the difference in average daily air pollution before and after the lockdown dates, using past years of data as controls. A closely related approach was taken by \citet{lee_chen_covid_2021}, who used synthetic controls in place of CITS. Similarly, \citet{rdit_ma} used regression discontinuity in time to estimate the air quality impact of a new public transport provision in London, at different sites.
However, these methods that model each location's effect separately do not satisfy our needs, because we are interested in characterizing the relationship between changes in air quality and each location's ``dose'' of exposure to the treatment / intervention, which in our case is port proximity as defined in Section~\ref{sec:port_prox}. That is, we wish to estimate a causal exposure-response curve.%

Our proposed method merges ideas from this literature on quasi-experiments with a ``universal'' intervention and ideas from the burgeoning literature on multiply-robust causal exposure-response curve estimation. Specifically, we build closely upon recent work by \citet{hettinger_MR}, who introduced a multiply-robust estimator for causal exposure-response curves within the difference-in-differences (DiD) framework. Derived from the efficient influence function or EIF (see Section \ref{sec:eif}), this estimator accounts for confounding both between treated and control units and across levels of exposure to the treatment / intervention, which can be thought of as a dosage. %
We propose a related EIF-based multiply-robust estimator for causal exposure-response curves in quasi-experimental designs with a universal intervention. We also propose an uncertainty quantification approach that accounts for both spatial and temporal correlation. We apply this method to remotely sensed data for the Indonesia plastic waste case study to estimate the change in air pollution at open dump sites due to increased plastic waste in the wake of China's 2018 ban, as a function of a dump site's port proximity.

\section{Data and Definitions}\label{data}

We compile a monthly analytic dataset spanning 2012-2019, not including more recent years to avoid confounding by the Covid-19 pandemic. This dataset will be made available on Harvard Dataverse post-acceptance of this paper.

\subsection{Spatial Units of Analysis: Waste Dump Sites}

Waste dump sites serve as our spatial units of analysis in this study. While plastic makes up only 12\% of global municipal solid waste, the total aggregation of waste is a key proxy for aggregations of plastic waste \citep{gpw_2023}. Indonesia does not provide official data on dump site locations in the country; thus, we source these data from remote sensing products. Global Plastic Watch (GPW) is a data platform containing satellite imagery-derived classifications of open dump sites, also known as waste aggregations. \citet{gpw_2023} used machine learning to do this classification, and evaluated their performance in Indonesia. The GPW dataset has been used by high-profile groups such as ClimateTrace to estimate emissions from the waste sector \citep{ClimateTrace_2023}. 

We use all dump sites in Indonesia detected by GPW prior to 2020 for which PM$_{2.5}$ data were available in the Asia region PM$_{2.5}$ data files from \cite{PM_data_2021}, described below. This results in 356 sites included in our analyses, out of 357 total detected by GPW in Indonesia.
Site footprints are provided monthly starting in 2015, but there are frequent gaps and GPW notes that the average across time is more reliable than any given time point. Averaged across time (through the end of 2019), the median area of open dump sites in Indonesia is 2,317m$^2$ (IQR=[881m$^2$, 5,982m$^2$]), but this distribution is highly right-skewed, with a maximum value of 288,424m$^2$ (0.29km$^2$). 

\subsubsection{Why Non-dump Sites Should Not Be Used as ``Control'' Locations}\label{controls} 

To motivate our methodological approach (in which we use past years as controls for future years at the GPW dump sites), briefly consider what trying to identify separate ``control" locations would look like. 
Let us walk through several of the causal identifying assumptions made in \cite{hettinger_MR}, which are standard in DiD-type analyses.

\textbf{Positivity} requires that controls have a non-zero probability of being treated, in other words, that non-dump sites selected as controls had some chance of receiving import-related plastic waste (which could either be imported waste or domestic waste displaced by imported waste) as a result of China's ban. In practice, we would need to be able to select non-dump locations with appropriate overlap of key confounders (e.g., population density and climate characteristics) such that they could be considered ``similar" locations to dump sites.
 
The \textbf{Stable Unit Treatment Value Assumption (SUTVA)} requires that (i) there is no interference between units and (ii) there are no hidden variations in treatment \citep{rubin1980randomization}. In our setting, spillover of air pollution becomes a concern if controls are spatially proximate to dump sites, which would be made more likely if we prioritized non-dump locations with similar landcover characteristics. The second aspect of SUTVA might be violated in our case study if we accidentally selected some ``control" locations where waste is dumped and burned, but dumping there went undetected in the satellite imagery. 

\textbf{Conditional Parallel Trends} requires that the pre-post difference in the treated group would be the same (conditional on covariates) as that in the control group, in the absence of treatment. Here, consider that Indonesia's domestic plastic waste generation is increasing by about 5\% per year \citep{NPAP_report}. Since we know that a substantial fraction of this waste is taken to dump sites and burned, we would expect the long-term trend in air quality over dump sites to be different than that over non-dump sites, regardless of waste being imported.

Taken together, these considerations illustrate why attempting to use non-dump sites as controls would likely lead to violations of causal identifying assumptions and biased estimates. Thus, we use only waste dump sites in our analysis and assume that they are universally treated by the China plastic waste ban starting in 2018. Note that we do not expect any dump sites to be sufficiently far away from ports to be unaffected by import-related plastic waste (such that they could be used as control locations) because Indonesia is an island nation and many plastic recycling facilities are known to be far inland \citep{nexus3_report}. Empirically, 98\% of GPW sites are within 100km of the coast; the farthest site is 150km inland. 

As an aside, it would be equally if not more problematic to try to use control locations in other countries because of differing (rates of change of) waste generation and burning practices, especially as many other countries in the East Asia \& Pacific region were similarly exposed to the effects of China's ban.

\subsection{Outcome: Fine Particulate Matter (PM$_{2.5}$) Concentrations}

Our outcome of interest is PM$_{2.5}$ concentrations at dump sites. While in-situ measurements are not available across Indonesia, \citet{PM_data_2021} estimated global monthly average PM$_{2.5}$ at 0.01° × 0.01° (approximately 1km) resolution using a combination of remotely sensed aerosol optical depth (AOD), chemical transport modeling, and in-situ measurements from around the world. Briefly, they relate AOD to surface PM$_{2.5}$ concentrations using geophysical relationships observed between in-situ PM$_{2.5}$ measurements and AOD simulated from a chemical transport model. They then apply geographically weighted regression to predict (and later adjust for) the residual bias with the in-situ monitors. At monitor sites in Asia from 2015-2019, they report cross-validated $R^2$ values ranging from 0.59--0.86. Unsurprisingly, estimates over regions with fewer in-situ monitors have higher uncertainty; however, this data is widely used for exposure assessment and epidemiology studies. 

We extract these monthly PM$_{2.5}$ values for 2012-2019 at the GPW sites, for each site taking the value of the grid cell in which the site centroid lies. The average PM$_{2.5}$ concentration at the dump sites in 2012-2017 (prior to China's ban) was 25.7 $\mu g/m^3$. 

\begin{figure}[h] %
    \centering
    \includegraphics[width=0.9\textwidth]{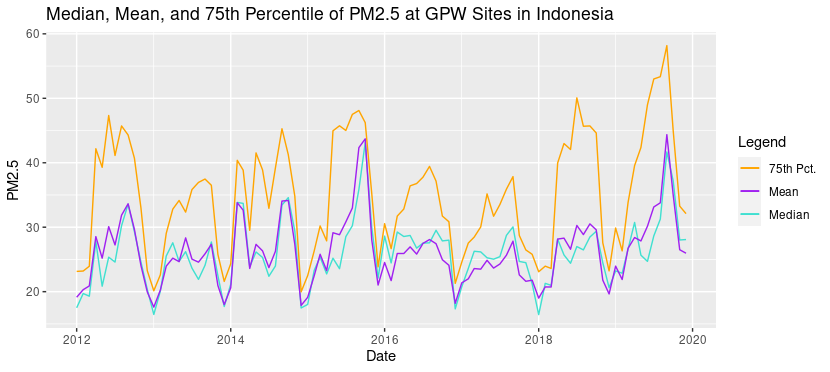} 
    \caption{Median, mean, and 75th percentile of PM$_{2.5}$ across GPW [dump] sites in Indonesia, 2012-2019.}
    \label{fig:overall_pm}  
\end{figure}

The time series of concentrations averaged across sites (Figure \ref{fig:overall_pm}) shows a strong seasonal pattern across our study period. However, without accounting for other time-varying factors, we cannot tell from a plot like this whether there was an unexpected increase in PM$_{2.5}$ in 2018-2019.

\subsection{Exposure: Port Proximity}\label{sec:port_prox}

\begin{figure}[h] %
    \centering
    \includegraphics[width=0.9\textwidth, trim={0, 60, 0, 60}, clip]{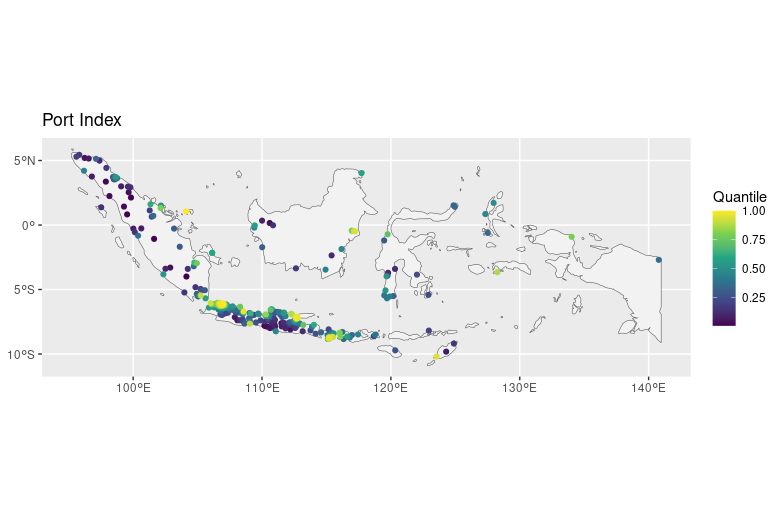} 
    \caption{Quantile of the port proximity index for each GPW site in Indonesia.}
    \label{fig:PPQ_map}  
\end{figure}

Because the policy intervention is on waste imports and in Indonesia any such imports should come through ports, presumably dump sites closer to port activity should be more heavily affected by waste imports. Thus, we consider each dump site's \textit{proximity to port activity} (described below) to be a proxy measure of treatment dose or ``exposure''. Note that to follow other recent work assessing the domestic impacts of China's plastic waste ban \citep{china_ban_AQ_2022}, we could use distance to the nearest port as a proxy; however, that would not take into account the amount of shipping activity in each location. Therefore, we used data from shipping signals remotely sensed via the Automated Identification System, specifically a derived measure of cargo ship loitering from the \citet{gmtds_loitering_2021}. For our continuous exposure metric, we calculated an inverse-distance-weighted sum of the 2018 loitering data around every GPW site, up to a threshold of 100km. Due to the strongly right-skewed distribution of this port proximity index, we use a quantile version throughout the analysis, as shown in Figure \ref{fig:PPQ_map}. This transformed variable is henceforth referred to simply as the ``port proximity".

\subsection{Covariates: Meteorology, Population Density, and Fire Locations}

As in previous analyses of pre/post-intervention air quality data, the primary covariates in our analyses are meteorological factors \citep{heffernan_2024,lee_chen_covid_2021,dey2021counterfactual}. By chance, meteorological conditions may differ in the pre- and post-intervention periods, also influencing air quality, and thus failure to adjust for them may bias our estimates of the treatment effects. We obtained ERA5 monthly-aggregated data for 2012-2019 \citep{era5_2017} via Google Earth Engine \citep{google_earth_engine_2017}. This climate reanalysis product, which integrates remotely-sensed and in-situ observations with physics-based models, %
is available at a spatial resolution of 0.25° × 0.25° (approximately 28km). We take the value at each dump site to be that of the ERA5 grid cell the dump site centroid lies within. We use average air temperature at 2m height, dewpoint temperature at 2m height (indicating humidity), total precipitation, surface pressure, and the u- and v-components of wind speed as covariates in our model. %

Population density is another potential confounder in our case study. The GPW API provides an estimate of the population living within 1, 5, and 10 kilometers of each dump site, sourced from WorldPop \citep{tatem_worldpop_2017}.
Because we observed similar patterns across the sites between these variables, we chose to adjust for population living within 1km for simplicity (it is straightforward to think of this as population density).

Lastly, we obtained the locations of active fires from VIIRS, the Visible Infrared Imaging Radiometer Suite \citep{viirs_active_fires}. The availability of this remotely sensed product (375m resolution daily) starting in 2012 influenced our selection of the study period start. A time series of the total number of active fires across Indonesia is shown in Figure \ref{fig:all_viirs}. 
After obtaining the maximum footprint of each dump site as detected by GPW, we performed a spatial overlay of the active fire points. To account for imprecise detection of site boundaries as well as shape fluctuation over time, we used a 100m buffer around each site for the overlay. Throughout our analysis, we consider non-waste dump fires to be those that occurred more than 100m away from any GPW site. 
We use the monthly count of fires occurring outside of open dump sites in the province as a covariate in our main analysis. The purpose of this is to adjust for potential changes over time in wildfire activity and agricultural burning practices, which could also contribute to changes in PM$_{2.5}$ during the study period.

To assess plausibility that changes in waste burning are contributing to any changes in air quality post-intervention, we conduct a complementary exploratory analysis of the temporal trend in the number of VIIRS active fire points at open dump sites in Indonesia during the study period. Increases in active fire points at open dump sites from 2018 onwards would provide informal evidence that changes in air quality may be (at least partially) due to increases in waste burning.

The VIIRS fire overlay also justifies our analysis of all dump site locations from 2012-2019, regardless of when GPW first detected them (prior to 2020). As shown in Figure \ref{fig:detection-date}, there were many locations at which major fires occurred prior to their date of first detection as a dump site by GPW. Having observed fires at a site before it was classified as a dump site by GPW increases the likelihood that the site was being used as a dump site prior to the GPW designation.

\section{Methods}\label{methods}

Analytic code and detailed information on data processing will be made available on GitHub post-acceptance of this paper. %

\subsection{Setup and Assumptions}\label{sec:setup} 
Let $\mathcal{O}$ be the observed data, containing: $Y$ the outcome (PM$_{2.5}$), $D$ the dose of exposure (port proximity), and $\mathbf{X}$ the covariates. The latter includes the population living within 1km of the site, monthly averaged meteorology (temperature, humidity, precipitation, pressure, and wind speed), and the number of active fires detected in the province (that month) which did not overlap with any GPW site. To help capture spatial heterogeneity, we also include fixed effects for site ID and province in our model; the spatial boundaries for the latter are shown in Figure \ref{fig:Indonesia_fire_map}. This is considered a reasonable approach in settings with many units where only few of these are important/informative \citep{fuhr_DML-panel_2024} and when both bigger-scale (e.g., province level) and smaller-scale (individual dump site level) heterogeneity are likely at play. Both of these characteristics are present in our case study, as shown by a complementary exploratory analysis (described in Section \ref{sec:eda}) illustrating that the distribution of the number of large fires (detectable by VIIRS) occurring at each dump site is highly skewed, with some clustering by province. Lastly, to account for longer-term trends, we include time $t=1,...,96$ to represent the months between 2012 and 2019, inclusive. Note that the seasonality illustrated in Figure \ref{fig:overall_pm} is encoded in the meteorological covariates and the number of active fires detected in the province which did not overlap with any GPW site, both of which are time-varying.

Furthermore, let $B$ (for ``ban") be a pre- vs. post-intervention indicator. $B=0$ indicates dates before policy implementation, and $B=1$ indicates dates after (in our case, 1/1/2018 onwards). Because all units (dump sites) in the analysis are treated post-intervention, $B$ is also a binary indicator of treatment status. We use $B$ rather than the more commonly used $A$ to illustrate this departure from DiD. As mentioned in the Introduction, while our use of time-varying covariates and pre-intervention years as controls for post-intervention years shares features with the synthetic control and interrupted time series literatures, existing methods do not enable estimating an exposure-response curve across locations.

Each observation is indexed by its location $i$ ($N$ = 356 dump sites) and time point $t$.  When $B_{it} = 0$, we say $D_{it} = \emptyset$, meaning $D$ is undefined -- while each location $i$ has a port proximity prior to China's ban, this feature has not yet been activated to represent the dose of treatment. We define the potential outcomes as $Y_{it}(B_{it}=b, D_{it}=\delta)$, i.e., the outcome that would have been observed for a given unit under intervention status $b$ and dose $\delta$. This definition of the potential outcomes requires that the future cannot affect the past (arrow of time, call this A1) and SUTVA (A2), discussed in Section \ref{controls}. Mathematically, SUTVA requires that $Y_{it}(\mathbf{B}_{t}, \mathbf{D}_{t}) = Y_{it}(B_{it}, D_{it})$, where $\mathbf{B}_{t}$ and $\mathbf{D}_{t}$ denote the population-level vectors of intervention and dosage. Note that while spillover of air pollution between dump sites would violate A2, this is unlikely to substantially impact our final effect curve because sites close together have similar "dose" values of port proximity, so our estimation procedure will average over them.

In order to avoid structural ill-definition, we also introduce an indicator for $t \geq T_0$, where $T_0$ is the treatment initiation time. In practice, $T_0$ is 1/1/2018 for all sites, but conceptually this could have occurred at any point in our study period, such that any time point could have been untreated. When $T_0$ is considered random, A5 (below) is a direct analog of the conventional conditional ignorability assumption used in cross-sectional data contexts, adapted to the time series setting.

Our aim is to estimate a causal exposure-response curve quantifying the causal effect on Indonesian air quality of China's 2018 plastic waste import ban at different levels of port proximity, $\delta$. In the DiD setting, \citet{hettinger_MR} terms this the ``Average Dose Effect on the Treated", or ADT. While noting again that our formulation is different, we still see fit to call our effect of interest the ADT: 
$$
ADT(\delta) = \Psi(\delta) := E[Y_{it}(1,\delta) - Y_{it}(0,\emptyset)|t \geq T_0]
$$

In our case study, this estimand describes the average difference in air quality post-China ban at open dump sites in Indonesia, if all dump sites were exposed to port proximity dose $\delta$. 

An important note about the ADT is how it differs from a more commonly-used Conditional Average Treatment Effect, $CATE(\delta) := E[Y_{it}(B=1) - Y_{it}(B=0)|D_{it}=\delta]$. First, recall that $D$ is undefined when $B=0$, not yet having been ‘activated’. Second, foreshadowing the following sections, the ADT approach enables interpreting the final effect curve \textit{in terms of port proximity} versus in terms of port proximity-possibly-confounded-by-other-factors. It does so by targeting confounding with respect to both $D$ and $B$, estimating a generalized propensity score for the former and a binary propensity score for the latter, in addition to modeling the outcome. A traditional doubly-robust estimator for the CATE would only use the binary propensity score.

In addition to A1 (arrow of time) and A2 (SUTVA), we assume 
$$
\text{(A3) Consistency: } Y_{it}(b,\delta) = Y_{it} \text{ when } (B_{it},D_{it})=(b,\delta)
$$

If $\mathbf{X}^B$ is the covariates excluding the time indicator $t$ and $\mathbf{X}^D$ is the covariates excluding both the dump site ID and the time indicator $t$, then we require

\begin{align*}
& \text{(A4) Positivity: there exists } \epsilon > 0 \text{ such that} \\
    & \:\:\:\: \text{(i) } \epsilon \leq \pi_B(\mathbf{x}^B) < 1 \: \forall \: \mathbf{x}^B \in \mathbb{X}^B, \:\:\:\:\:\:\:\: \pi_B(\mathbf{x}^B) = P(t \geq T_0|\mathbf{x}^B) \\
    & \:\:\:\: \text{(ii) } \epsilon \leq \pi_D(\delta|\mathbf{x}^D,B=1) \: \forall \: \mathbf{x}^D \in \mathbb{X}^D|B=1, \delta \in \mathbb{D}, \:\:\:\: \pi_D(\delta|\mathbf{x^D},B=1) = p(D=\delta|B=1, \mathbf{X}^D=\mathbf{x}^D) \\
\end{align*}

While $P(t \geq T_0|\mathbf{x}^B)$ does not represent a feasible data generating process, this formulation highlights that there must not be separation in the covariate space between the pre- and post-intervention time periods. Flexibly estimating $\pi_B$ characterizes the extent of this overlap, rather than attempting to capture a true data generating process.

In contrast to Hettinger's Conditional Counterfactual Parallel Trends Between Treated and Control, we require
$$
\text{(A5) Conditional Ignorability Between Time Periods: } 
Y_{it}(0,\emptyset) \perp\!\!\!\perp I(t \geq T_0) \: | \: \mathbf{X}_{it}
$$

This means that there is \textit{no unmeasured time-varying confounding} -- no unmeasured factor causing air pollution to be systematically different in the periods before and after China’s ban, after accounting for confounding factors. Assembling an adequate set of covariates to justify this assumption relies on domain knowledge. Its plausibility can be strengthened by adjusting for seasonal and/or long-term trends, e.g., through inclusion of a time variable in the adjustment set.  
Note that while China's ban being enacted should not affect confounders like meteorology and population density in Indonesia, we must pay careful attention to any other policies with the potential to make air pollution, conditional on covariates, systematically different in the $B=0$ and $B=1$ time periods -- see discussion of this in Section \ref{sec:discussion}.

Lastly, in contrast to Hettinger's Conditional Counterfactual Parallel Trends Among Treated Between Doses, we require
\begin{align*}
& \text{(A6) Conditional Ignorability Among Treated Between Doses: } \\ & 
Y_{it}(1, \delta) \perp\!\!\!\perp D_{it} \: | \: t \geq T_0, \mathbf{X}_{it} \: \: \: \: \: \: \: \: \forall \: \: \delta \in \mathbb{D}
\end{align*}
This means that there is \textit{no unmeasured spatial confounding}, in other words we would expect the post-policy air quality to be the same at all dump sites with covariates $\mathbf{X}$, if they were assigned dose $\delta$ (regardless of their observed port proximity). The plausibility of this assumption may be strengthened by adjusting for region-specific fixed effects (in our case, province).

Conditional ignorability (in both A5 and A6) is a stronger assumption than parallel trends. Any method for treatment effect evaluation in the absence of a control group must make assumptions to rule out unforecastable shocks that affect all outcomes in the post-treatment period \citep{botosaru_forecasted}. Similarly, for a continuous treatment, while parallel trends would adjust for baseline differences between units receiving different doses, we must assume that our covariate set is able to explain these baseline differences. For our application, the alternative of assuming that some locations represent the control condition post-treatment (China's ban) would likely be inaccurate, as described in Section \ref{controls}. 

Note that while we believe it is reasonable to assume conditional ignorability on distributions here given our domain knowledge, all that is needed for identifiability of our estimand is (a weaker) conditional ignorability in expectations, as is shown below.

Assuming these six conditions, our $ADT(\delta)$ is identifiable with observed data:
\begin{align}
\Psi(\delta) &= E[Y(1,\delta) - Y(0,\emptyset)|t \geq T_0] \\
    &= E[E[Y(1,\delta) - Y(0,\emptyset) |t \geq T_0, \mathbf{X}] | t \geq T_0] \\
    &= E[E[Y(1, \delta) | t \geq T_0, D=\delta, \mathbf{X}] - E[Y(0,\emptyset)|t < T_0, \mathbf{X}] | t \geq T_0] \\
    &= E[E[Y |B=1, D=\delta, \mathbf{X}] - E[Y|B=0, \mathbf{X}] | B=1] 
\end{align}

Where line (1) is by definition, (2) by iterated expectations, (3) by A6 on the lefthand side and A5 on the righthand side, and (4) by definition of $B$ and A1-3.

Finally, note that our port proximity index, $D$, is a proxy of the true continuous exposure variable of interest, i.e. the amount of China ban-instigated import-related plastic waste, which could either be imported waste or domestic waste displaced by imported waste.  (Figure \ref{fig:dag} illustrates this with a directed acyclic graph.) We denote this true exposure $G$. Thus, our use of $D_{it}$ as a proxy for $G_{it}$ requires

\begin{align*}
& \text{(A7) Valid Proxy for the Continuous Exposure: $D$ and $G$ satisfy both of the following } \\ 
    & \:\:\:\: \text{(i) }  D_{it} \not\!\perp\!\!\!\perp G_{it} \\
    & \:\:\:\: \text{(ii) }  Y_{it} \perp\!\!\!\perp D_{it} \: | \: G_{it}, X_{it} \\
\end{align*}

\subsection{Motivations for Efficient Influence Function-Based Estimation}\label{sec:eif}

A conventional approach to estimate an exposure-response curve in a quasi-experimental setting would be to use a regression model like this one: 
$$
E[Y_{it}] = \beta_0 + \boldsymbol{\beta} \mathbf{X}_{it} + \tau B_t + f_{\phi}(D_i)B_t
$$
where $f_{\phi}(D)$ is a flexible (nonparametric) function, with parameters $\phi$. Note that a simplified version of this, where $f_{\phi}(D) = \phi D$, has been extensively analyzed -- for instance see \citet{callaway_2024}. This approach has several drawbacks. To begin, even if we allowed a flexible relationship between $D$ and $Y$, we would not be allowing for flexible relationships between $\mathbf{X}$ and $Y$, nor interactions between the components of $\mathbf{X}$. Importantly, we would not be addressing confounding between the dose $D$ and trends in the outcome -- for instance, if the overall (or seasonal) trend in air pollution is affected by population density, which is static in our dataset and which we know to be correlated with $D$ (see Section \ref{sec:eda}). To address this, we could make a much more complicated linear regression (e.g., by using splines and more interaction terms), or use a nonparametric machine learning algorithm to model $Y$ given $D,\mathbf{X},B$. However, this outcome model might still be misspecified.
This motivates the search for a doubly-robust estimator, which incorporates both a model for the outcome and a model for the propensity score (the probability of receiving treatment), which explicitly addresses confounding, and will produce results that are consistent (reliable in a probabilistic sense) if at least one of these models is correctly specified. %

We turn to estimators based on the the Efficient Influence Function (EIF). Without going into the technical details (interested readers should see \citet{kennedy2022} and \citet{modern_causal_schuler}), the EIF for a specific estimand can be used to construct an estimator that is doubly robust and attains the nonparametric efficiency bound (a nonparametric analog of the Cramer-Rao bound), at least asymptotically. Note that, as we will see below, the EIF does not always construct a \textit{doubly}-robust estimator \citep{modern_causal_schuler}, and the EIF may not exist for a particular estimand. In practice, one has to derive the EIF only for a novel estimand, and in many cases this can build closely on past work. 
\citet{kennedy2017} provided a doubly-robust estimation method for causal exposure-response curves in observational data settings, which \citet{hettinger_MR} then extended to a DiD framework, including conditions for ``multiple robustness" which will be described in the following section. We further extend this multiply-robust estimator to a setting without control locations, which as described in Section \ref{intro} is another common type of quasi-experiment. This work contributes to the nascent application of EIF-based estimation techniques in quasi-experimental settings.

\subsection{Estimation}\label{sec:estimation}

The major difference between our estimator and that of \citet{hettinger_MR} is that instead of modeling the pre-post difference in outcomes,
we are just modeling the outcome $Y$ because we are using pre-intervention years rather than separate control locations. This does not substantively change the steps of the proofs in \citet{hettinger_MR} to (1) derive the EIF and (2) show that the resulting estimator is multiply robust. A technical note is that the derivation of the EIF here refers to the integral of $ADT(\delta)$ to satisfy the pathwise differentiability requirement.

At a high level, after obtaining the EIF, we construct the estimator via the method of estimating equations, which involves setting the EIF equal to zero and solving for $\Psi$ \citep{modern_causal_schuler}. Given our goal of estimating $\Psi(\delta)$, the relevant parts of this estimator are: 

\begin{align*}
& \xi(D, \mathbf{X}, B, Y; \mu_1, \pi_D) = B*\left[\frac{f(D|B=1)}{\pi_D(D|\mathbf{X}^D, B=1)}(Y-\mu_1(D,\mathbf{X})) + m(D|B=1)\right] \\
    & \tau(\mathbf{X}, B, Y; \mu_0, \pi_B) = \frac{1}{P(B=1)}\left[\frac{(1-B)*\pi_B(\mathbf{X}^B)}{(1 - \pi_B(\mathbf{X}^B))}(Y-\mu_0(\mathbf{X})) + B*\mu_0(\mathbf{X}) \right] \\
\end{align*}
where
\begin{align*}
    & \mu_0(\mathbf{X}) = E[Y|B=0,\mathbf{X}] \\ 
    & \mu_1(D,\mathbf{X}) = E[Y|B=1, D, \mathbf{X}] \\
    & m(D|B=1) = \int_\mathbb{X} \mu_1(D,\mathbf{x}) dP(\mathbf{x}|B=1) = E[Y|B=1,D] \\
    & f(D|B=1) = \int_\mathbb{X}\pi_D(D|B=1, \mathbf{x}^D)dP(\mathbf{x}^D|B=1) \\
\end{align*}

In words: $\mu_0$ and  $\mu_1$ are the outcome regressions in the pre- and post-intervention periods, respectively, and $m(D|B=1)$ is the marginalized outcome regression in the post-intervention (treated) period, averaged across the covariates so that it only depends on the dose of exposure $D$. $\pi_B$ is the propensity score for being intervened upon (in our case, being in the treated period), $\pi_D$ is the generalized propensity score (GPS) for receiving dose $D$ of treatment, given being in the treated period, and $f(D|B=1)$ is the marginalized GPS.  

Overall, this estimator is unusual in that it has separate functions $\xi$ and $\tau$ to estimate the outcome in the presence and absence of treatment -- in our case meaning the years after intervention and before intervention respectively. The form of both functions, including both an outcome model and a propensity score-based weight, foreshadows the multiple robustness conditions of this estimator: we need at least one of \textit{each pair} $(\mu_1, \pi_D)$ and $(\mu_0, \pi_B)$ to be correctly specified for the results to be consistent. 

To estimate $\Psi(\delta)$, we follow this procedure:

\begin{enumerate}
    \item Estimate nuisance functions $\mu_1, \mu_0, \pi_B, \pi_D$
    \item Calculate $\hat{\xi}(D, \mathbf{X}, B, Y; \hat{\mu}_1, \hat{\pi}_D)$ and $\hat{\tau}(\mathbf{X}, B, Y; \hat{\mu}_0, \hat{\pi}_B)$ on the empirical data using the estimated nuisance functions
    \item Regress (smooth) $\hat{\xi}$ on $D$ to obtain $\hat{\Psi}_D(\delta)$, and take the empirical mean of $\hat{\tau}$ to obtain $\hat{\Psi}_0$
    \item $\hat{\Psi}(\delta) = \hat{\Psi}_D(\delta) - \hat{\Psi}_0$
\end{enumerate}

The estimation of the nuisance functions can be done with flexible modeling techniques subject to some constraints which prevent overfitting. For multiple robustness, $\mu_1$ and $\pi_D$ must either be fit using methods from uniformly bounded function classes with finite uniform entropy integrals (referred to as the Donsker class) or use cross-fitting / sample splitting, where the functions are estimated on one part of the data and used to predict in a held-out fold. The methods used to fit $\mu_0$ and $\pi_B$ just have to converge in probability to the truth (as $n \rightarrow \infty$).
Recent work on estimating treatment effects in the absence of a control group \citep{botosaru_forecasted} emphasizes the importance of unbiasedness rather than predictive accuracy when forecasting the counterfactual for the treated based on pre-treatment observations. (Other aspects of \cite{botosaru_forecasted} do not apply in our setting because we have a much longer time series and complex covariate space that cannot easily be forecasted with basis functions.)

Local linear kernel regression (LLKR) is employed for the smooth regression in step 3. The methods and implementation details used in our case study analysis are described in Section~\ref{sec:sensitivity_methods}.

\subsection{Uncertainty Quantification} 

For uncertainty quantification, \citet{hettinger_MR} derive a sandwich variance estimator, but mention various limitations of that approach, including that it relies on a normal approximation, that incorporating uncertainty in nuisance function estimation is challenging, and that extensions would be required to account for correlated data. To avoid these complications, they use a nonparametric bootstrap approach to obtain pointwise confidence intervals for the ADT. Rather than discrete bootstrapping, where observations are either included in a sample or left out entirely, they use weights drawn from an Exponential distribution to allow for small but nonzero inclusion, which helps to improve the observed support of the continuous exposure (in our notation, $D$) in the samples \citep{hettinger_MR}. To account for temporal correlation across repeated observations, they sample one weight per location (as opposed to per individual observation). These weights are scaled to reflect the observed sample sizes in each treatment group.

An important extension we propose, motivated by our case study, is to adapt the weights to account for spatial correlation in addition to temporal. We also modify the weights-scaling step to reflect the observed sample sizes in the pre- and post-intervention time periods, as we do not have separate control locations under a ``universal" intervention.

Note that our approach to addressing spatial correlation is distinct from the weighted block bootstrap approach in a working paper by \citet{hettinger_framework}, in which they sample a weight per block/neighborhood defined as all adjacent ZIP codes to a focal ZIP code, and then sum the weights assigned within each ZIP code from overlapping blocks.
Described below, our approach has the advantages of (a) empirically estimating the residual spatial correlation and (b) avoiding artifacts/inconsistencies stemming from the use of location-areas with irregular sizes and/or shapes, such as two sites $m$ miles apart being considered spatially correlated in some places but not in other places, and ZIP codes with more adjacencies (shared borders) systematically receiving larger weights.

\subsubsection{Spatial Weighted Bootstrap}

Because our set of covariates likely does not explain all spatial correlation in the data, we need to account for residual spatial dependence in our uncertainty quantification. For instance, note that there are some clusters of dump sites near Jakarta (shown in Figure \ref{fig:Jakarta-fire-map}) which we would be hard-pressed to call independent of one another, even conditional on our covariate set. Therefore, in our weighted bootstrap procedure we would like a site's weight to be larger (or smaller) if other sites nearby have larger (or smaller) weights. We achieve this through \textit{spatially-correlated sampling of the bootstrap weights.}

To build intuition, imagine the following procedure to generate the weights, given a spatial correlation matrix $\Sigma$:
\begin{enumerate}
    \item Sample from a Multivariate Normal distribution with mean zero and variance-covariance (in this case, correlation) matrix $\Sigma$. 
    \item Apply the inverse probability transform\footnote{Also known as the inverse transformation method, this is a basic technique to generate a sample from any probability distribution given its cumulative distribution function or CDF $F_X(x)$. In brief, if $U \sim \text{Uniform}[0,1]$, then we can generate $X = F_X^{-1}(U)$. If we start by sampling $Z$ from a distribution with CDF $F_Z(z)$, then we can generate $X = F_X^{-1}(F_Z(Z))$.} to obtain samples with a different (target) marginal distribution, such as the Exponential.
\end{enumerate}

In practice, we can estimate $\Sigma$ by modeling the empirical spatial variogram of the residuals (see the next section for more details on this estimation and explanation of which model's residuals are used). This approach of generating bootstrap weights from an estimated spatial correlation function is substantiated by \citet{SDWB_2024}, who show the mathematical validity of what they term the spatially dependent wild bootstrap (SDWB). While the SDWB paper assumes a Gaussian random field, they note that this assumption can be relaxed at the expense of additional technical complexity. The issue, described in depth by \citet{anySim}, is that the magnitude of correlation is not preserved during the mapping from Gaussian random variables to those from another distribution using the inverse probability transform. To end up with the desired spatial correlation in our Exponential weights, we must first estimate an ``equivalent correlation" $\Sigma^*$ to feed into the Multivariate Normal sampling procedure. Namely, $\Sigma^*$ should be specified such that, when the Multivariate Normal samples are inverse probability transformed to generate Exponential weights, the resulting weights have the desired spatial correlation ($\Sigma$). The estimation of $\Sigma^*$ can be achieved using numerical integration, which is facilitated by the \texttt{anySim} package \citep{anySim} in R. 

The complete proposed Spatial Weighted Bootstrap procedure is detailed below.
\begin{enumerate}
    \item Estimate the empirical variogram of the model residuals and use it to estimate the desired spatial correlation matrix for the weights, $\Sigma$ (see next section).
    \item Estimate the equivalent correlation matrix $\Sigma^*$ for the desired marginal distribution of the weights (in our case, Exponential). 
    \item Sample weights $\tilde{w}_i$ (one per site) from a Multivariate Normal distribution with mean zero and variance-covariance matrix $\Sigma^*$. 
    \item Apply the inverse probability transform to $\tilde{w}_i$ to obtain weights $w_i$ with a marginal Exponential distribution and correlation matrix $\Sigma$. 
    \item Use $w_i$ as weights for each observation from location $i$ in the dataset ($t=1,...,96$), and carry these weights through nuisance function estimation and regression on $D$ to obtain an estimate of the ADT.
    \item Repeat steps 3-5 to generate the bootstrap estimates of the ADT. 
\end{enumerate}

Point-wise confidence intervals for $ADT(\delta)$ can be formed by extracting the appropriate percentiles (in our case, 2.5 and 97.5) of the estimates across the bootstrap samples.

\subsubsection{Estimating the Residual Spatial Correlation Function}\label{sec:residual-corr}

In conventional outcome-model-only settings, residual spatial correlation refers to spatial autocorrelation in the residuals obtained from subtracting the model-predicted outcome values $\hat{Y}$ from the observed values $Y$. In our setting, the precise definition of residual spatial correlation is less evident due to the multiple stages of modeling that go into our estimation: both outcome and propensity score models are plugged into the EIF-derived equations $\xi$ and $\tau$, then $\hat{\xi}$ is regressed on $D$. 

For convenience, let $\hat{\xi}_{it}$ denote $\hat{\xi}(D_{it}, X_{it}, B_{it}, Y_{it})$. Here, we focus on spatial correlation in the residuals $\hat{\xi}_{it} - \hat{\Psi}_D(D_{it})$ for all observations in the treated period $B=1$. (Note that the counterfactual outcome $\hat{\Psi}_0$ is constant over space.)
Intuitively, we address residual spatial correlation from the final stage of modeling because that is the stage in which our effect of interest is estimated, and it represents the ``last opportunity'' to explain spatial correlation in the data via modeling. Note that in settings like ours where the exposure of interest is spatially explicit, calculating residuals based on the final effect curve (which has been marginalized over all the other covariates) induces a different spatial structure than if we were to analyze residuals $Y_{it} - \hat{\xi}_{it}$. 
From a more technical perspective, \citet{Lee_2017_DR_confidence_band} note that for doubly-robust methods utilizing LLKR, the model-based uncertainty arising from the second-stage estimation (the LLKR) is greater than that of the first-stage estimation (the nuisance models). Having ``no estimation effect" from the first stage is closely related to the conditions for double (or multiple) robustness, constraining the nuisance functions as mentioned in Section \ref{sec:estimation}.

After calculating the residuals $\hat{\xi}_{it} - \hat{\Psi}_D(D_{it})$, we estimate the empirical variogram using the robust estimator introduced by \citet{cressie_1980_robust}, which is less susceptible to influential outliers than the basic method-of-moments approach. We then fit a covariance function, $Cov(m; L, \sigma^2)$, to the empirical variogram, where $m$ is the distance between each pair of sites, $L$ is a bandwidth or ``range" parameter, and $\sigma^2$ is the magnitude of the variance. The corresponding correlation function $Corr(m; L)$ is obtained by removing the constant $\sigma^2$.

\subsubsection{Simulations to Investigate the Spatial Weighted Bootstrap}

A pseudo-simulation study was conducted to evaluate the properties of the proposed Spatial Weighted Bootstrap confidence intervals. In Appendix \ref{sec:simulations}, we provide full details on the data generating process (based closely on our real-world dataset), the calculation of performance metrics, and the results of this simulation study.

These simulations illustrate that the Spatial Weighted Bootstrap has consistently higher coverage than the Non-Spatial Weighted Bootstrap, due in part to generating wider confidence intervals. Averaging across all the values of the exposure (port proximity), the Spatial approach has a coverage of 96.5\% and the Non-Spatial approach has a coverage of 91.8\%. However, both methods tend to systematically have higher-than-nominal coverage in the middle of the exposure distribution and lower coverage on the tails. Across several simulation scenarios, the lowest coverage observed (at any value of the exposure) for the Spatial Weighted Bootstrap is 87\% whereas for the Non-Spatial approach it dips down to 82\% (see Figure \ref{fig:coverage}). Re-evaluating the coverage after subtracting out the average bias illustrates that some, but not all, of the under-coverage is attributable to the finite sample bias of the LLKR (see Figure \ref{fig:coverage-debiased}).

\subsection{Case Study Data Application}\label{sec:sensitivity_methods}

In our case study analysis, we use the Ranger algorithm in the \texttt{caret} package \citep{caret_package} in R to estimate the nuisance functions, using separate models for $\mu_1, \mu_0, \pi_B, \pi_D$. 
Ranger is an efficient implementation of random forest, which has been shown to work well in tabular data settings like this \citep{tabular_tree_based}, and also implicitly performs a type of variable selection \citep{islr_textbook}, which is helpful for identifying which of the fixed effects, at the province or individual site level, is informative \citep{tabular_tree_based}.
However, because this algorithm does not conform to the finite uniform entropy integral condition used to prove multiple robustness, we use cross-fitting with k=10 sample splits for $\mu_1$ and $\pi_D$. (For $\mu_0$ and $\pi_B$ we do not do cross-fitting to reduce the computational intensity -- as it is not needed for multiple robustness -- instead carefully tuning hyperparameters as described below. For applications with simpler covariate spaces and/or shorter pre-treatment time series, a parametric or semi-parametric model to forecast the counterfactual may be preferred \citep{botosaru_forecasted}.) Note that Donsker class alternatives like Highly Adaptive Lasso (HAL) tend to be substantially more computationally intensive than Ranger, even with cross-fitting. 
Another important note is that cross-fitting in a spatiotemporal setting like this one is an active area of research. Splitting samples into folds by unit and by time period both have their disadvantages. Interestingly, in a variety of panel data simulations, \citet{fuhr_DML-panel_2024} found violations of the cross-fitting independence assumption to be largely inconsequential for the accuracy of the effect estimates.
 
Recall from Section \ref{sec:setup} that the set of covariates used in the two propensity score models excludes the time (month) indicator $t$. For $\pi_B$, this is because the set of covariates in a flexible machine learning model must not be able to perfectly predict the intervention \citep{heffernan_2024} -- in our case, time steps before and after 1/1/2018. For $\pi_D$, this is because our port proximity values are static within the $B=1$ time period.
Additionally, the covariates used to estimate $\pi_D$ do not include the dump site ID because the port proximity value is unique to each site. Because of this, we sample observations by site into the cross-fitting folds (in addition to initial hyperparameter tuning).

We carefully tune each Ranger model according to its use in our estimation procedure. 
For the regression models (all except for $\pi_B$, which is a binary classification), we use min.node.size=5 and select mtry (the number of variables to possibly split at in each node) using 5-fold cross-validation. For $\mu_1$, we sample observations into these folds randomly (within $\mathcal{O}_{B=1}$) and set mtry to be the value beyond which the $R^2$ and RMSE do not substantially improve (a difference within 0.01 for $R^2$ and 0.1 for RMSE). For $\pi_D$, we instead sample observations by site into the folds to account for repeated observations at each location. For $\pi_B$, we sample observations randomly from $\mathcal{O}$ and limit mtry such that the empirical probabilities do not get too large ($>$0.99) or too small ($<$0.001). Lastly, for $\mu_0$, we set the folds to be 2012-2014 vs 2015-2017 to emulate this model's extrapolation to the $B=1$ years, and select mtry to minimize the bias rather than maximize predictive accuracy, as recommended in \citet{botosaru_forecasted}. 
For all the models, we use num.trees=500.

To regress $\hat{\xi}$ on $D$ (step 3 in the estimation procedure from Section~\ref{sec:estimation}), we use the \texttt{locpol} package \citep{locpol_package} in R to perform LLKR with a Gaussian kernel. This package utilizes leave-one-out cross-validation to select the bandwidth parameter for the kernel regression. Therefore, naively applying the method to our dataset, which contains repeated observations (over time) at each unique value of $D$ (one per dump site), selects an inappropriately small bandwidth.
Therefore, we modified this package's bandwidth selection function to do leave-one-\textit{location}-out cross-validation where all repeated observations for a given location are assigned to the same fold, as has been done in other kernel regression analyses with longitudinal data \citep{Kernel_BW_CV}.

For the Spatial Weighted Bootstrap, we compared several standard spatial covariance functions (Exponential, Linear, Spherical, and Mat\'ern), ultimately selecting the Mat\'ern as the one with the best RMSE when fit to the empirical variogram of the residuals from the LLKR stage. %
Therefore, the entries of $\Sigma$ for our spatially-correlated weights have the form 

$$
Corr(m; L, \nu) = \frac{2^{1-\nu} }{\Gamma(\nu)} \left( \frac{m}{L} \right)^\nu K_\nu \left( \frac{m}{L} \right)
$$

where $\Gamma$ is the gamma function, $K_\nu$ is the modified Bessel function of the second kind, and $\nu$ is an additional smoothness parameter which is optimized alongside $L$.

\begin{figure}[!ht] 
    \centering
    \includegraphics[width=0.4\textwidth]{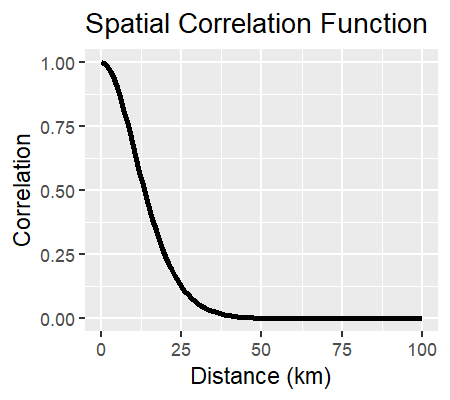} 
    \caption{The spatial correlation function (Mat\'ern) fit to the residuals from the LLKR stage; $\hat{L} = 3.85$km and $\hat{\nu}=5$.}
    \label{fig:spat-corr}  
\end{figure}

The empirical spatial range was $\hat{L} = 3.85$km, and the optimal smoothness parameter was $\hat{\nu}=5$. Figure \ref{fig:spat-corr} illustrates the resulting magnitude of correlation for sites at different distances apart.

Confidence intervals for the ADT (with or without spatial correlation in the weights) are calculated using 100 bootstrap samples. In our simulation study, increasing the number of bootstrap replicates to 200 did not substantially change the confidence intervals' coverage, as shown in Figure \ref{fig:num_boots}.

For illustrative purposes, we compare the results from our multiply-robust EIF-based method to (i) a plain outcome regression, which is the same as that used inside the multiply-robust estimator, followed by regression on $D$ using LLKR, and (ii) a flexible but confounder-naive approach. The latter estimates $E[Y|D=\delta, B=1] - E[Y|B=0]$, where the first expectation uses LLKR to regress on $D$ and the second expectation is a sample mean. 

We also re-ran the analysis using 2017 and, separately, 2016 \& 2017 as the ``treated" period, to check for pre-trends / policy anticipation (whether the effect began before 2018). Conducting such tests is standard practice for quasi-experiments \citep{bilinski_pretest_2026, roth_pretest_2022}; \citet{heffernan_2024} demonstrated this technique while utilizing past years as controls, in other words, utilizing a placebo in time.  
Note that this kind of pre-trends / placebo testing is technically assessing the null hypothesis of no pre-trends; for more formal evaluation, it is preferred to test whether the magnitude of the pre-trend effect is sufficiently small as to not invalidate the estimated treatment effect \citep{bilinski_pretest_2026}. Future work might explore development of such tests for settings like ours.

This pre-trends validation approach connects to our stated assumptions in several ways. Clearly, anticipation of a policy being implemented would violate A1 (arrow of time). 
If the pre-trends analysis finds an effect of treatment in 2017 (prior to treatment initiation), this could also indicate that some important time-varying factors influencing PM$_{2.5}$ are not being captured by our measured confounders, in violation of A5. (A policy implemented in 2017 could fit this criteria, as noted in our Discussion.) However, the absence of an effect in the pre-trends test only offers evidence of plausibility of A5 but does not formally validate it because we cannot test for unmeasured time-varying confounding between $B=0$ and $B=1$, instead relying on domain knowledge. Also, note that because our pre-trends test is estimating the ADT, it is inherently testing conditional ignorability in expectation. 

Lastly, we conducted a sensitivity analysis to assess the potential influence of dump site size, as it could be an important predictor of air pollution heterogeneity across sites. As mentioned in Section \ref{data}, Global Plastic Watch estimates the area (footprint) of each dump site. Although time-varying estimates are provided, these measurements are less reliable than their average over time and have frequent gaps, so we used each site's average area across available estimates (2012-2019).
We did not adjust for dump site size in the main analysis because area measured after 1/1/18 is a post-treatment variable which can be affected by treatment, referred to as a ``descendant" of treatment, which should not be conditioned on in causal inference analyses \citep{causal_what_if}. Note that only 68\% of the dump sites have at least one area estimate prior to 2018, so we would have to exclude many sites, potentially biasing our causal effect curve, in order to include only pre-2018 area as a covariate.

For simplicity, this sensitivity analysis as well as the aforementioned outcome regression, confounder-naive approach, and pre-trends tests use the same nuisance function hyperparameters and Spatial Weighted Bootstrap with the same Mat\'ern parameters as our main analysis.

\section{Results}

\subsection{Statistical Analysis Results}

\begin{figure}[!h] 
    \centering
    \includegraphics[width=0.9\textwidth]{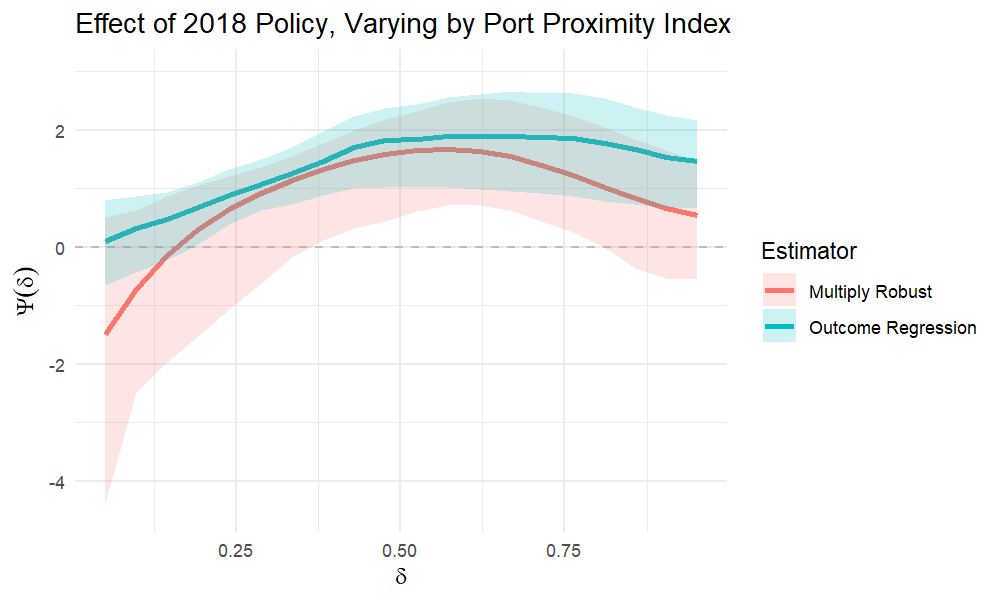} 
    \caption{Average dose-effect on the treated (ADT) estimates and bootstrap 95\% confidence intervals for our case study, estimated using both our proposed multiply robust method and a conventional outcome regression. ADT estimates (y-axis) quantify the average change in PM$_{2.5}$ ($\mu g/m^3$) post-China ban, compared to concentrations expected under business-as-usual, at open dump sites in Indonesia for a given port proximity (x-axis). Note that our port proximity metric is in quantile form, so the observed support of the exposure along the x-axis of these plots is constant. The pointwise confidence intervals are generated using our Spatial Weighted Bootstrap.}
    \label{fig:EIF_main}  
\end{figure}

Figure \ref{fig:EIF_main} shows the estimated ADT curve and 95\% bootstrap confidence interval (CI) from the main analysis. Positive ADT estimates provide evidence of post-China ban increases in PM$_{2.5}$ concentrations at GPW sites, compared to those expected under business-as-usual. Thus, our results indicate that at open dump sites in Indonesia, the monthly PM$_{2.5}$ in 2018-2019 was generally higher than expected under business-as-usual (except for dump sites nowhere near port activity), with the effect getting as large as 1.68 $\mu g/ m^3$ (95\% CI = [0.72, 2.48]) depending on the port proximity. As expected, we see smaller effects with lower port proximity -- the pointwise confidence intervals include zero until $\delta=0.38$. Interestingly, the effect curve peaks in the middle and decreases for high values of port proximity, including zero again at $\delta=0.81$ -- our intuition is that this could reflect smaller increases in dumping/burning in densely populated areas very close to ports, where there is likely more government oversight and less open space.
Integrating across all values of the estimated effect curve yields an average increase at dump sites of 0.86 $\mu g/m^3$ attributable to China's ban on plastic imports, corresponding to a 3.3\% increase in overall PM$_{2.5}$ in 2018-2019 compared to 2012-2017. 
For illustration of the heterogeneity: if we only consider values of port proximity for which the confidence intervals do not include zero, the average increase is 1.51 $\mu g/m^3$. 

Similarity between the effect curve from our proposed multiply-robust method and that produced using an outcome regression alone (Figure \ref{fig:EIF_main}) suggests a fairly strong outcome model. Differences between these two methods at the tails may indicate that the generalized propensity score is doing its job (guarding against misspecification of the outcome model) or may reflect LLKR having instability at the extremes, as seen in our simulations (Appendix \ref{sec:simulations}).
For a more extreme comparison, the confounder-naive results are shown in Figure \ref{EIF_Naive-only}: the mean effect ranges from -6.4 to 8.5 $\mu g/m^3$. This divergence from our main result is unsurprising, but demonstrative.

The results of ignoring residual spatial correlation in the bootstrap stage are shown in Figure \ref{fig:grid-compare-spatial-ranges}. The confidence intervals are, on average, 73\% the width of the intervals obtained from Spatial Weighted Bootstrap. Note that this is for our multiply robust method, which also incorporates spatial-folds (all observations from a site sampled together) cross-fitting when estimating the generalized propensity score. For the outcome regression alone, the confidence intervals from the Non-Spatial Weighted Bootstrap are 52\% the width of the intervals from the Spatial Weighted Bootstrap, on average.
This demonstrates the importance of accounting for residual spatial correlation.

\begin{figure}[!h] 
    \centering
    \includegraphics[width=0.9\textwidth]{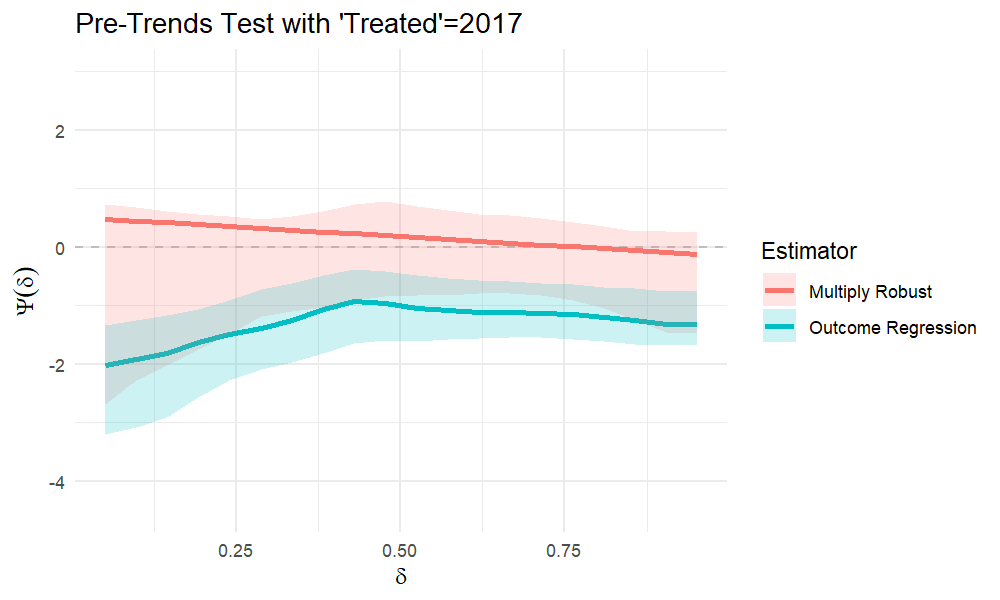} 
    \caption{Pre-trends test using 2017 as the ``treated" period; using Spatial Weighted Bootstrap with the same Mat\'ern parameters as our main analysis.}
    \label{fig:EIF_A-2017}  
\end{figure}

As shown in Figure \ref{fig:EIF_A-2017}, we did not observe an effect when using 2017 as the ``treated" period (to check for pre-trends / policy anticipation). Using 2016 \& 2017 as the ``treated" period (Figure \ref{fig:EIF_A-2016-2017}) similarly shows no evidence of anticipation.

Lastly, including GPW site area as a covariate in our sensitivity analysis (Figure \ref{fig:with_site_area}) did not substantially change the results.

\subsection{Complementary Exploratory Data Analysis Results}\label{sec:eda}

\begin{figure}[h] %
    \centering
    \includegraphics[width=0.9\textwidth]{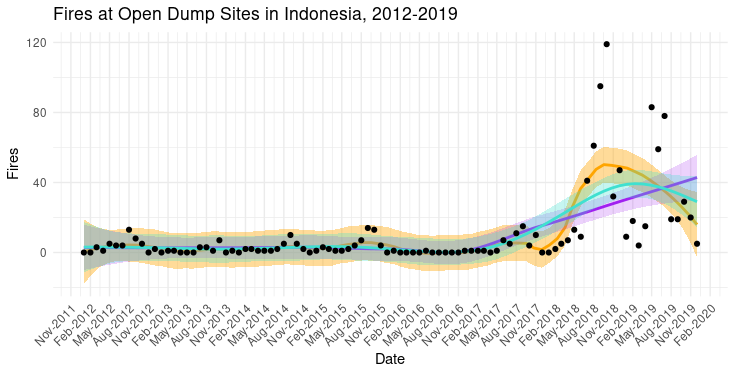} 
    \caption{The number of fires detected by VIIRS overlapping with GPW sites (within a 100m buffer) across Indonesia. The colored trend lines are estimated using several different smoothing techniques: the orange using LOESS with span=0.25, the purple using LOESS with span=0.75, and the turquoise using a generalized additive model.}
    \label{fig:VIIRS_time_series}  
\end{figure}

To complement our main analysis, we investigated the occurrence of active fires at the GPW dump sites throughout our study period. Figure \ref{fig:VIIRS_time_series} shows the time series (aggregated by month) of VIIRS-detected fires overlapping with the dump sites. We observe a large increase in the number of fires in 2018-2019.  

The distribution of the number of fires detected by VIIRS at each dump site is highly skewed, as shown in Figures \ref{fig:Indonesia_fire_map} and \ref{fig:Jakarta-fire-map} (maps of Indonesia overall and the Jakarta region respectively). This is somewhat explained by the area of the dump sites, but not entirely (Figure \ref{fig:fires-area}). In keeping with our main analysis, the sites with the most fires are not those with the highest port proximity (Figure \ref{fig:fires-ppi}) -- there is a spike around $\delta=0.7$, due to the cluster of large dump sites to the west of Jakarta, which partially motivated our Spatial Weighted Bootstrap approach. 
We also note that the correlation between the port proximity and (square root-transformed) population living within 1km of a site is 0.6. While this collinearity might be an issue for estimating our effect of interest in a linear model, it is less of a concern for a doubly-robust nonparametric estimator in which we marginalize over the covariates.

\section{Discussion}\label{sec:discussion}

In this case study, we observed a 3.3\% average increase in PM$_{2.5}$ at dump sites in Indonesia in 2018-2019 compared to 2012-2017, attributable to China's plastic waste import ban. 
This effect is on a similar scale to the results of \citet{china_ban_AQ_2022}, who assessed the domestic impacts of China's ban and estimated an average reduction of 3.7 $\mu g/m^3$, corresponding to 8.5\% reduction of overall PM$_{2.5}$ concentrations in China as a result of their 2018 ban. It makes sense that our observed effect would be smaller, given that Indonesia was only one of the countries to which waste that would otherwise have gone to China was diverted. \citet{china_ban_AQ_2022} similarly did not observe an anticipation effect in mid-2017, when China announced its upcoming ban to the World Trade Organization. 

Methodologically, we demonstrate the viability of pairing spatiotemporal causal inference methods with remotely sensed data products to quantify the air quality impacts of large-scale plastic waste policies via the mechanism of waste burning. On the data side, we harmonize areal classifications of open dump sites, cargo ship loitering densities, meteorologic estimates, active fire points, and high-resolution air pollution estimates. For the statistical analysis, we combine a multiply-robust
estimator for causal exposure-response curves within the difference-in-differences framework with the technique of using pre-intervention years as controls for post-intervention years. Additionally, we illustrate the importance of accounting for spatial dependence in the uncertainty quantification stage, and propose a Spatial Weighted Bootstrap approach for doing so. Given that there are numerous documented quasi-experiments created by plastic waste policies globally \citep{nicholas_plastics_policy_inventory}, which have largely occurred in countries lacking in-situ air quality monitoring data, we anticipate that the analytic framework developed here can be much more widely applied.

Beyond the setting of plastic waste and air quality, this paper offers two important tools that span a wide range of quasi-experimental studies: (1) a robust method to estimate the impact of a ``universal" intervention (such that there are no control locations) as a function of a continuous exposure and (2) a generalizable inferential procedure for spatially correlated data in the context of a continuous exposure. 

In the interest of helping practitioners understand these tools, we note that interpretation / inference for an effect curve like ours is inherently more complicated than for the effect of a binary treatment: we are interested in both the magnitude and the shape of the curve, to assess ``compatibility" \citep{rafi_compatibility_2020} with our hypothesis that air pollution in Indonesia increased in the wake of China's 2018 ban, heterogeneously with respect to port proximity. A more nuanced aspect is that the plausibility of the curve shape influences our interpretation of the magnitude.
In this analysis, our findings of no effect at low values of port proximity, a substantial effect at medium-high values of port proximity, and a weaker/suggestive effect at very-high values of port proximity are very plausible and compatible with our domain knowledge.
To be clear, dose values whose pointwise confidence interval for the ADT excludes zero (the null) may be interpreted as doses for which there is some evidence of a treatment effect; however, the use of \textit{uniform} confidence bands may be preferred for formal inferential purposes to avoid multiple comparisons concerns or to test hypotheses regarding the functional form of the effect curve. Hettinger et al. 2025a note in their Web Appendix F that uniform confidence intervals could potentially be obtained using a combination of cross-fitting and scaling of the bootstrapped standard errors by a constant specific to both the kernel and the bandwidth parameter used in LLKR and the range of the dose variable, which is an extension of \citet{Lee_2017_DR_confidence_band}. However, as mentioned in our Section \ref{sec:sensitivity_methods}, cross-fitting in a spatiotemporal setting like ours is an active area of research.

This study has a few limitations which merit remarks.
First, while we are focused on the air quality impact of China's 2018 ban and resultant increase in plastic waste in Indonesia, the observed difference between 2018-2019 and 2012-2017 might be affected by another policy that was implemented around the same time (2017-2018): Indonesia's National Action Plan on Marine Plastic Debris. The goal of this policy was to reduce the amount of plastic waste ending up in the ocean -- with limited emphasis on reducing the amount of plastic waste being generated \citep{marine_policy_2023}; by 2021 Indonesia achieved an estimated 15\% reduction of marine plastic debris from 2017 levels \citep{nicholas_institute_report}.
Ethnographic studies from around the world have found that campaigns to raise awareness about plastic pollution often contribute to increased open burning, as communities seek to cut down on litter and deal with waste that has been collected \citep{global_health_2024}. Hence, it is possible that these two policies were interacting to generate the observed increase in PM$_{2.5}$ at dump sites in Indonesia. Our analysis showing no pre-trends effect when using 2017 as the ``treated" year guards against this possibility somewhat, however the effect of the National Action Plan could lag or increase after 2017. 

Second, while our interest is primarily on air pollution generated by the burning of waste, it is possible that some of the observed increase in PM$_{2.5}$ at open dump sites is attributable to the \textit{transportation} of this waste, for instance from diesel trucks making more trips to dump sites and, for sites close to the coast, from cargo ships.
Our complementary (exploratory) analysis illustrating the uptick in active fires overlapping with dump sites offers some evidence that at least part of the observed effect is due to increases in waste burning. Unfortunately, speciated PM$_{2.5}$ estimates, which might enable the disentangling of emissions from waste burning and transportation, tend to be available only at coarse spatiotemporal resolutions globally and rely more heavily on models \citep{edgar_v8}, which is likely to frustrate the application of quasi-experimental statistical methods. For smaller scale studies, \citet{bangladesh_plastic_AQ_2022} demonstrated the viability of measuring a molecular tracer of plastic burning with in-situ monitors. Alternatively, 
\citet{maldives_ML_2023} applied machine learning-based image classification to identify smoke plumes in high-resolution (3m x 3m) satellite imagery over the largest municipal landfill in the Maldives. However, in addition to being computationally intensive, this approach is not directly translatable into air pollution concentrations, which are easier to link to the rich literature on health impacts, and hence tend to be more relevant for policy makers.

Third, while remotely sensed data products enable application of quasi-experimental methods in data-scarce settings, we must be conscious of their potential for measurement error / misclassification. In this case study, we believe that the GPW dump site classifications are the data product in which errors would be most consequential for our analysis. Because all the GPW sites detected in Indonesia were manually validated \citep{gpw_2023}, there should be no false positive classifications. However, false negative classifications are possible and could affect our analysis in two ways. On the one hand, because we include the number of active fires in each province \textit{not overlapping with dump sites} as a covariate in $\mathbf{X}$, any fires mistakenly classified as non-dump fires will reduce our estimated policy effect. On the other hand, if there is spillover of PM$_{2.5}$ from false negative dump sites to nearby GPW-detected dump sites, then this could increase our estimated \textit{per-site} policy effect. 

A more general limitation of our approach is that we cannot estimate the effects of distributed waste burning, i.e., at the household level or at communal dump sites that were too small to be detected by GPW. Future research, and likely investment in ground-level air quality monitors (and/or lower-cost sensors), will be needed to estimate these distributed burning emissions, which may ultimately pose greater risks to public health due to higher frequency of burning, greater proximity to people, and lower dispersive dilution of air pollution at the ground level \citep{global_health_2024} compared to taller plumes of smoke resulting from larger fires at larger dump sites. 

Despite its limitations, this study provides evidence that the export of plastic waste to Indonesia, and subsequent open burning of import-related waste in Indonesia, resulted in higher concentrations of PM$_{2.5}$ -- at least near large open dump sites. This adds to the bodies of work documenting (a) air quality impacts of plastic waste and (b) environmental degradation tied to the global waste trade.
Policy interventions to address open burning of plastic waste will have to contend with the phenomenon that simply prohibiting open burning is seldom effective \citep{global_health_2024}, though burning at major sites may be easier to monitor and restrict \citep{maldives_ML_2023}. 
In terms of the plastic waste trade: in 2021, Indonesia restricted the import of non-hazardous waste to 15 specific ports \citep{nexus3_report} and, in 2025, banned their import of plastic waste altogether \citep{indonesia_ban_2025}. In mid-2025, Malaysia followed suit, only allowing plastic waste from parties that have ratified the Basel Convention on the Control of Transboundary Movements of Hazardous Wastes and their Disposal \citep{malaysia_ban_2025}. 
The effectiveness of these and future policies, at least for improving air quality but potentially also for other kinds of environmental outcomes, can in turn be evaluated using methods similar to those presented in this paper. %

\newpage

\section{Competing interests}
No competing interest is declared.

\section{Author contributions statement}

E.M.C. conceptualized this project and compiled the dataset. E.M.C. and R.C.N. developed the statistical method. E.M.C. wrote the code and carried out the analysis. E.M.C. drafted the manuscript, and both E.M.C. and R.C.N. edited the manuscript. E.M.C. led the revisions in response to reviewer comments, with feedback from R.C.N.

\section{Acknowledgements}

This work was supported by an NSF Graduate Research
Fellowship (EMC), NIH award K01ES032458 (RCN), and the Harvard Climate Change Solutions Fund (RCN).
The computation for this paper was performed on (a) the FASRC Cannon cluster supported by the FAS Division of Science Research Computing at Harvard University and (b) the Alpine high performance computing resource at the University of Colorado Boulder. Alpine is jointly funded by the University of Colorado Boulder, the University of Colorado Anschutz, and Colorado State University and with support from NSF grants OAC-2201538 and OAC-2322260.

We would like to thank the National Studies on Air Pollution and Health research group for their feedback and support. Additionally, we thank Christine Wiedinmyer and Randall Martin for early input on the project direction, and the Nafas team for sharing their knowledge of the air quality monitoring, evaluation, and regulatory landscape in Indonesia.

\clearpage
\bibliographystyle{abbrvnat}
\bibliography{references}

\clearpage

\begin{appendix}

\renewcommand{\thefigure}{S\arabic{figure}}
\renewcommand{\thetable}{S\arabic{table}}
\setcounter{figure}{0}
\setcounter{table}{0}

\section{Estimation Details}\label{estimation-details}

\subsection{Generalized Propensity Score (GPS) Estimation}

Following the GPS estimation method utilized by \citet{causalGPS_2024}, we first fit a model to $D$ given $\mathbf{X}$, within $B=1$. Note that while in our dataset, $D$ is static (derived from the 2018 GMTDS data), early experimentation revealed that the time-varying covariates are meaningfully predictive of $D$. Our intuition for this is that the spatial variation in the \textit{average} values of the meteorological covariates, such as humidity, is informative of how close one is to the coast, which is in turn associated with port proximity.

After fitting the initial model for $D$, we fit a model to the residuals $D_i - \hat{D}_i$ given $\mathbf{X}$, within $B=1$. (For efficiency, we use the same hyperparameters for this residual model as for the initial model of $D$.) Then, we apply a kernel density estimator to $(D_i - \hat{D}_i)/\hat{\epsilon}_i$ to obtain $\pi_D$. For $f(D|B=1) = \int_\mathbb{X}\pi_D(D|B=1, \mathbf{x})dP(\mathbf{x}|B=1)$, we interpolate over the observed $\mathbf{X}$ for specified values of $D$.

\section{Supplementary Figures and Tables}

\begin{figure}[h] %
    \centering
    \includegraphics[width=0.9\textwidth]{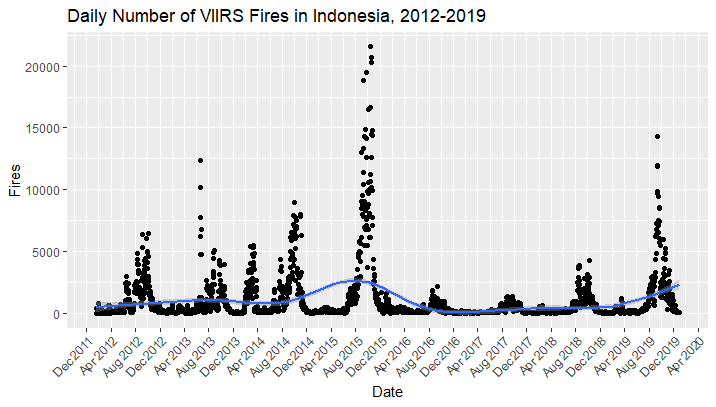} 
    \caption{Sum of daily VIIRS fires detected anywhere in Indonesia (not only at waste dump sites), with smoothed trend shown.}
    \label{fig:all_viirs}  
\end{figure}

\begin{figure}[h] %
    \centering
    \includegraphics[width=0.9\textwidth]{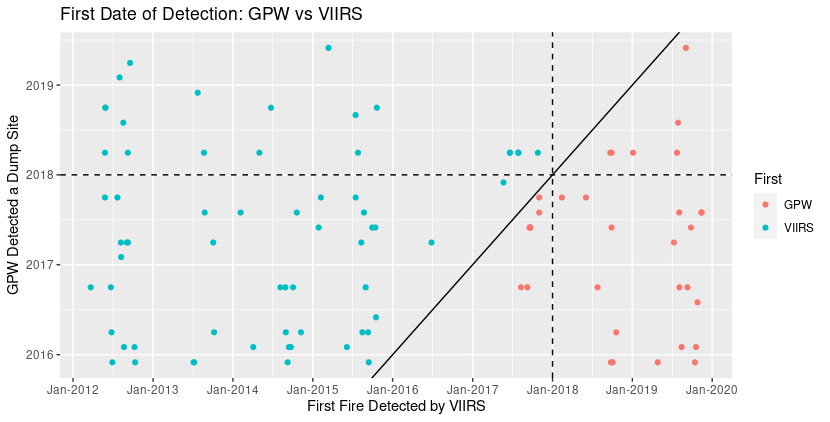} 
    \caption{Detection date of each dump site (at which active fires occurred) based on GPW's algorithm and VIIRS.}
    \label{fig:detection-date}  
\end{figure}

\begin{figure}[h] %
    \centering
    \includegraphics[width=1.0\textwidth]{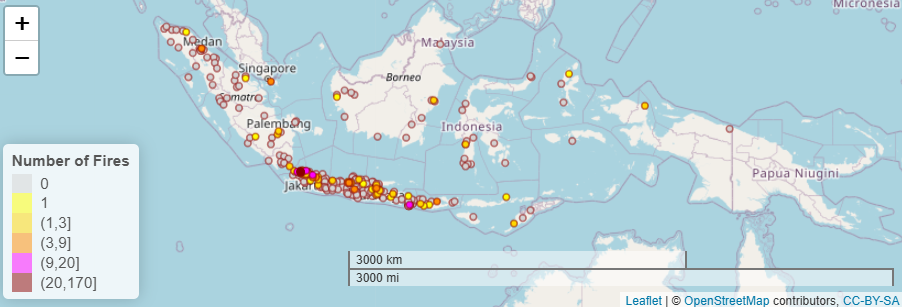} 
    \caption{Map of Indonesia showing the number of fires detected by VIIRS overlapping with each dump site (detected by Global Plastic Watch) in 2012-2019 %
    and the boundaries of the provinces (in gray).}
    \label{fig:Indonesia_fire_map}  
\end{figure}

\begin{figure}[h] %
    \centering
    \includegraphics[width=0.9\textwidth]{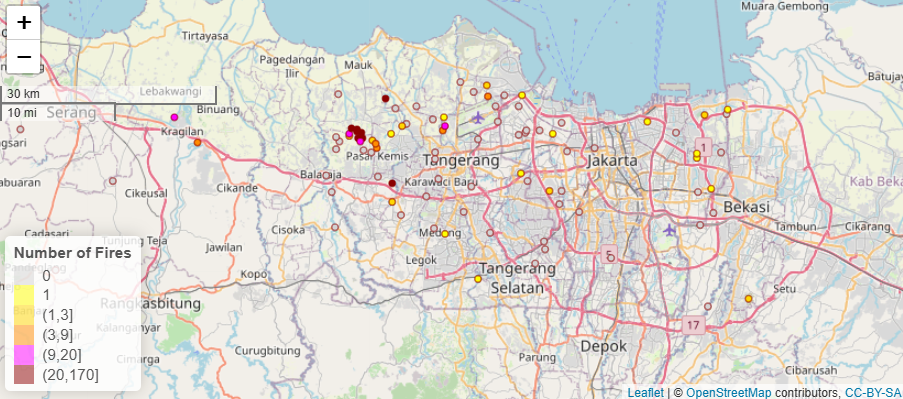} 
    \caption{Zoomed-in map of Jakarta showing the number of fires detected by VIIRS overlapping with each GPW site in 2012-2019%
    .}
    \label{fig:Jakarta-fire-map}  
\end{figure}

\begin{figure}[h]
    \centering
    \includegraphics[width=1.0\textwidth]{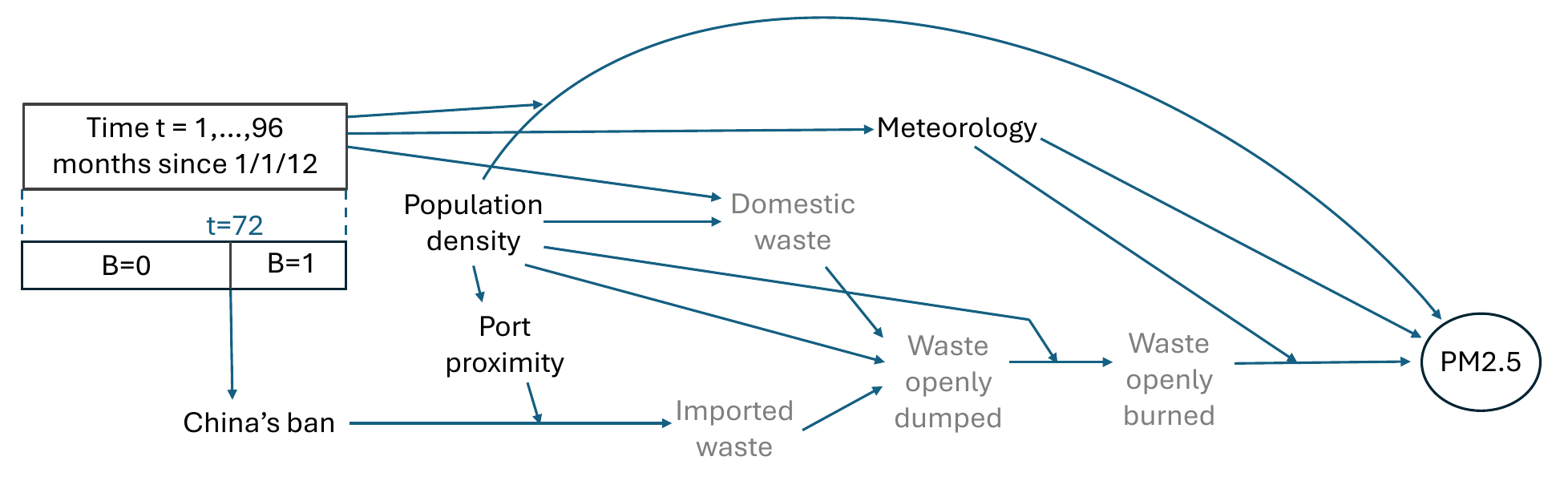}
    \caption{Directed Acyclic Graph (DAG) illustrating our application. Unobserved variables are shown in gray, while observed variables are shown in black. Interactions are suggested by intersecting arrows, as proposed by \citet{dag_effectmod}. Note that this is a slightly simplified DAG, intended to motivate our use of port proximity, rather than all the remotely sensed proxies we utilize.}
    \label{fig:dag}
\end{figure}

\begin{figure}[!h] 
    \centering
    \includegraphics[width=0.8\textwidth]{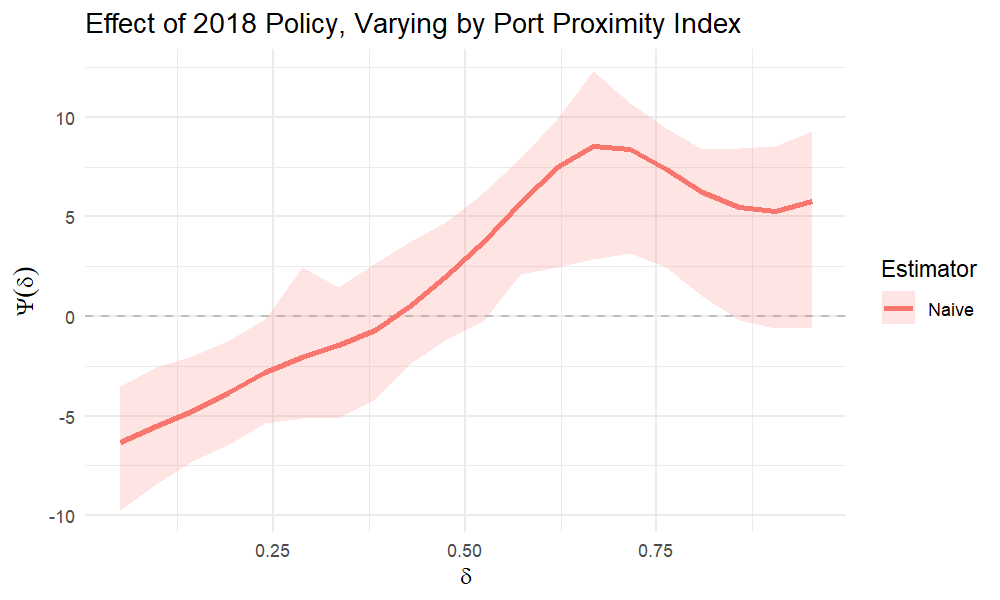} 
    \caption{The confounder-naive results; using Spatial Weighted Bootstrap with the same Mat\'ern parameters as our main analysis.}
    \label{EIF_Naive-only}  
\end{figure}

\begin{figure}[!h] 
    \centering
    \includegraphics[width=1\textwidth]{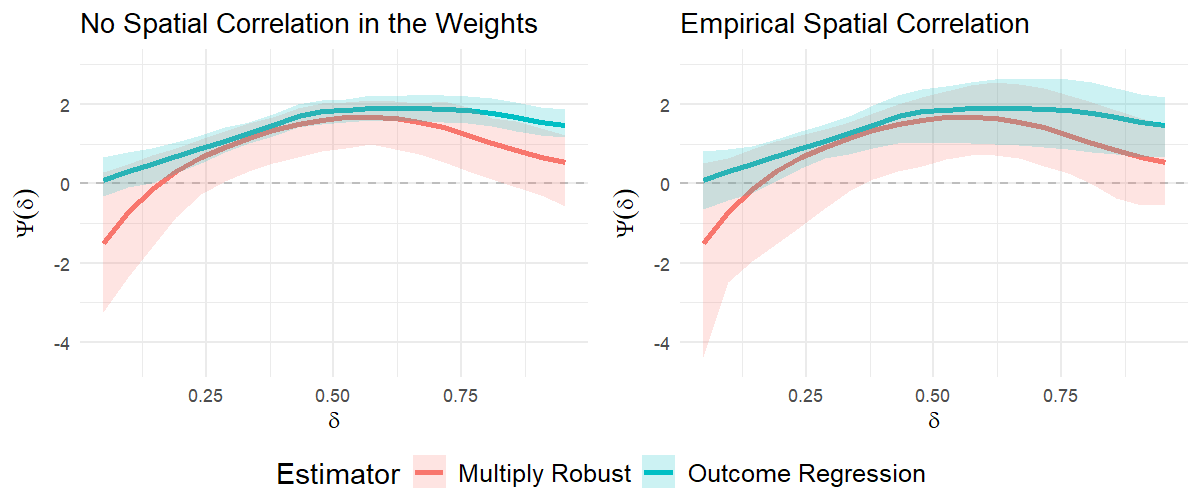} 
    \caption{Comparing the results using Non-Spatial and Spatial Weighted Bootstrap. Note that the point estimates are the same.}
    \label{fig:grid-compare-spatial-ranges}  
\end{figure}

\begin{figure}[!h] 
    \centering
    \includegraphics[width=0.9\textwidth]{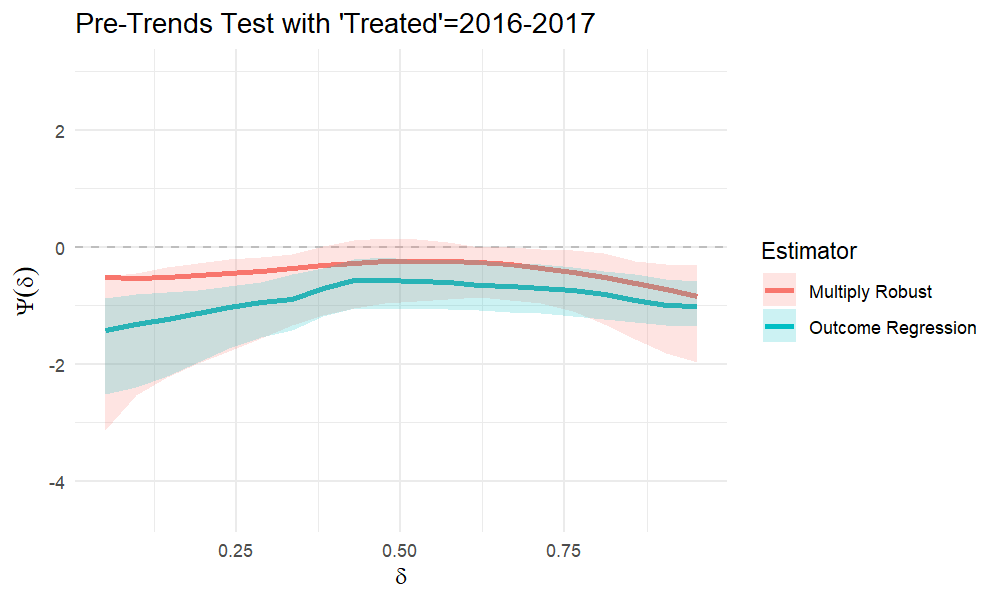} 
    \caption{Pre-trends test for 2016-2017; using Spatial Weighted Bootstrap with the same Mat\'ern parameters as our main analysis.}
    \label{fig:EIF_A-2016-2017}  
\end{figure} 

\begin{figure}[!h] 
    \centering
    \includegraphics[width=0.9\textwidth]{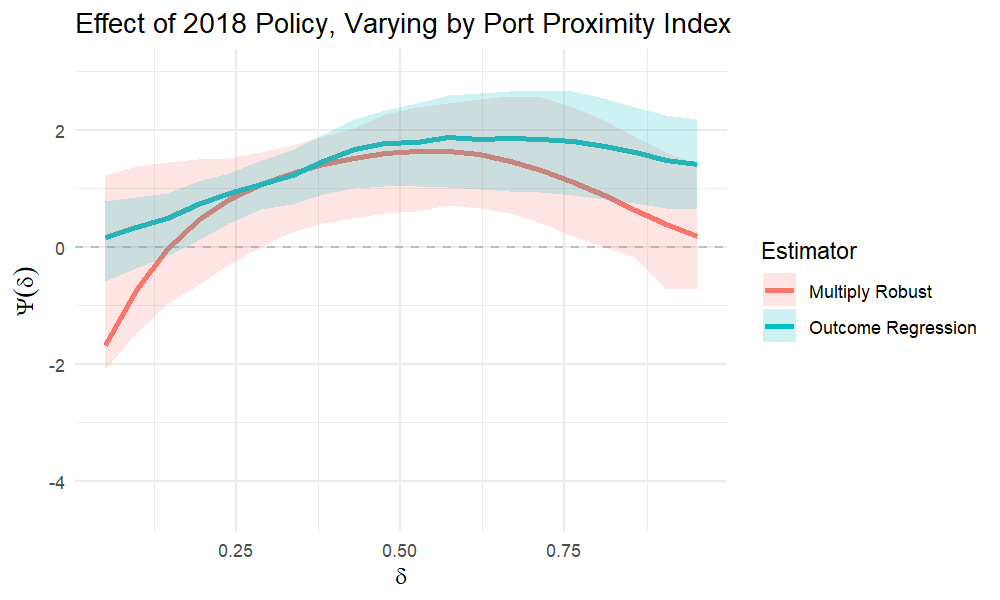} 
    \caption{Sensitivity analysis including GPW site area as a covariate; using Spatial Weighted Bootstrap with the same Mat\'ern parameters as our main analysis.}
    \label{fig:with_site_area}  
\end{figure}

\begin{figure}[h] %
    \centering
    \includegraphics[width=0.8\textwidth]{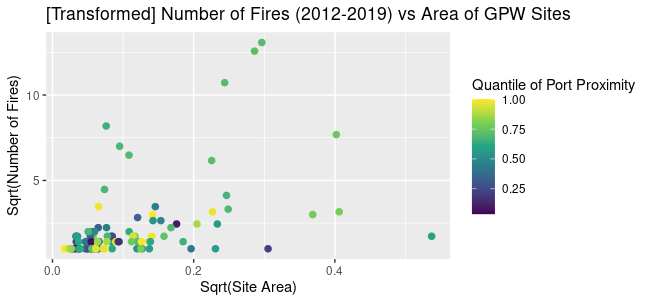} 
    \caption{Square root of number of fires detected by VIIRS vs square root of GPW site area, colored by the quantile of port proximity.}
    \label{fig:fires-area}  
\end{figure}

\begin{figure}[h] %
    \centering
    \includegraphics[width=0.8\textwidth]{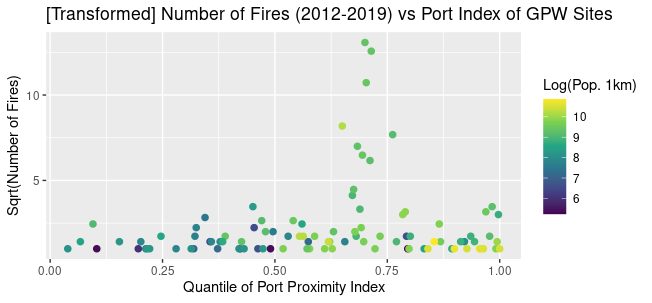} 
    \caption{Square root of number of fires detected by VIIRS vs quantile of port proximity, colored by the log of population living within 1km of each GPW site.}
    \label{fig:fires-ppi}  
\end{figure}

\clearpage

\section{Simulations Exploring the Spatial Weighted Bootstrap}\label{sec:simulations}

In this appendix, our primary aim is to evaluate the characteristics of the confidence intervals generated by our Spatial Weighted Bootstrap procedure, relative to those generated by a Non-spatial Weighted Bootstrap. Here, calculation of the ADT curve's bias and mean absolute error (MAE) is an intermediate step to better understand the confidence intervals, as \citet{hettinger_MR} present extensive investigation of a very similar ADT estimator. 

To do this evaluation, we develop simulations which utilize the real (observed) exposure $D$ and covariates $\mathbf{X}$ to simulate the outcome $Y$.

\subsection{Simulating the Outcome}

First, we compute the mean function for the outcome data generating process using the real exposure and covariate data and setting ``true" parameter values governing the relationship between these variables and the outcome in our simulated data. Our choice of these parameter values was guided by (a) using domain knowledge to select the sign of each coefficient, (b) producing an outcome distribution similar to that in the real data, and (c) producing a plausible ADT curve, similar to that observed in our main analysis.

The mean function for each location $i$ and time $t$ is given by:
$$ 
\mu_{it} = B_{t} L_1(\mathbf{X}_{it}, D_{it}) + (1-B_{t})L_0(\mathbf{X}_{it})
$$

where

\begin{align*}
L_0(\mathbf{X}_{it}) &= -9.5 + 0.075(\text{Temp}_{it}) + 0.075(\text{Pop}_{i}) + 0.0001(\text{Temp}_{it})(\text{Pop}_{i}) + 0.15(\text{Time}_{t}) +\\
   & \:\:\:\:\:\: \:\:\:\:\:\: 
   0.001(\text{Time}_{t})(\text{Pop}_{i}) +
    1(\text{Other\_fires\_province}_{it}) - 18(\text{Precip}_{it})
\end{align*}

and

\begin{align*}
L_1(\mathbf{X}_{it}, D_{it}) &= -12.5 + 0.075(\text{Temp}_{it}) + 0.075(\text{Pop}_{i}) + 0.0007(\text{Temp}_{it})(\text{Pop}_{i}) + 0.15(\text{Time}_{t}) +\\
   & \:\:\:\:\:\: \:\:\:\:\:\: 
   0.001(\text{Time}_{t})(\text{Pop}_{i}) +
    1.5(\text{Other\_fires\_province}_{it}) - 20(\text{Precip}_{it}) + \\
    & \:\:\:\:\:\: \:\:\:\:\:\: 
    2(D_{it})\Big[ -3(D_{it}) + 0.4(\text{Pop}_{i}) + 0.2(\text{Time}_{t})\Big]
\end{align*}

where ``Temp" is temperature, ``Pop" is population [density], ``Time" is the fifth-root of the count of days since 01/01/2012, ``Other\_fires\_province" is the fourth-root of the number of fires in the province not overlapping with any GPW site, and ``Precip" is precipitation. 
For both $L_0$ and $L_1$, we truncated any negative values ($\sim$0.06\%) to zero post hoc. 

We calculate the ``true" ADT for each exposure value $\delta$ 
as 
$$
\Psi(\delta) = \frac{1}{96N} \sum_{i=1}^N \sum_{t=1}^{96} \
\Big[ L_1(\mathbf{X}_{it}, \delta) - L_0(\mathbf{X}_{it}) \Big]
$$

\begin{figure}[ht] 
    \centering
    \includegraphics[width=0.9\textwidth]{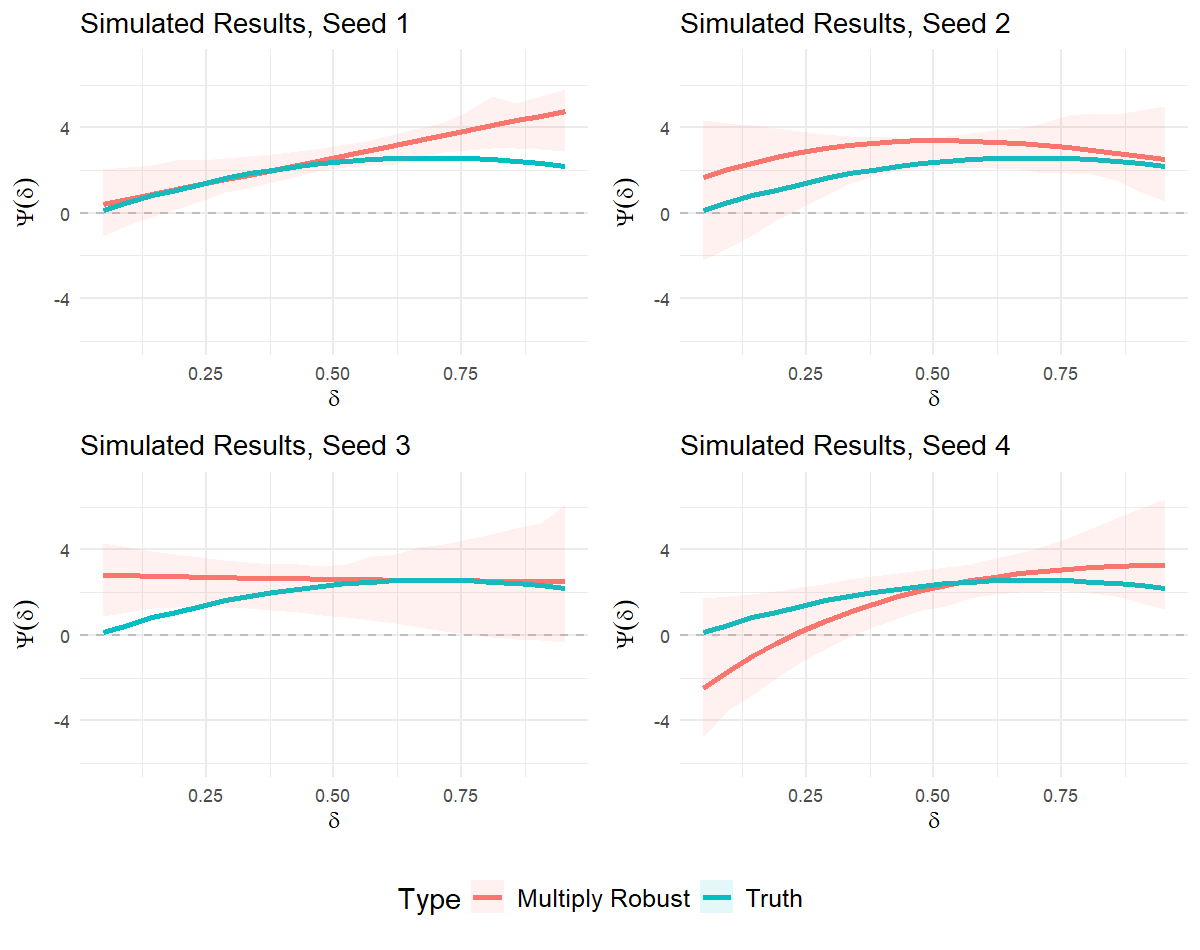} 
    \caption{Examples of the ADT estimate from our multiply robust estimation procedure applied to (four instances of) simulated data from our base case scenario, compared to the true ADT curve.}
    \label{fig:sim-example}  
\end{figure}

This true ADT curve is shown in Figure \ref{fig:sim-example}.

We simulate both spatially-structured and independent error in the outcome, as follows:
$$
\tilde{Y}_{it} = \mu_{it} + \sigma_\text{sp} \gamma_i + \sigma_\text{ind} \epsilon_{it}
$$

where $\gamma_i \sim MVN(0, \Sigma)$, where the correlation function to generate the entries of $\Sigma$ is $Corr(m;L) = exp(-m/L)$ [Exponential form] and $\epsilon_{it} \sim N(0, 1)$. The parameters $L$ (the true spatial range), $\sigma_\text{sp}$, and $\sigma_\text{ind}$ are constants, which we specify to obtain four distinct simulation scenarios:
\begin{enumerate}
    \item Scenario 1: $L=100$ km, $\sigma_\text{sp} = 2$, $\sigma_\text{ind} = 0.5$ (the ``base case")
    \item Scenario 2: $L=500$ km, $\sigma_\text{sp} = 2$, $\sigma_\text{ind} = 0.5$ (increased spatial range)
    \item Scenario 3: $L=100$ km, $\sigma_\text{sp} = 3$, $\sigma_\text{ind} = 0.5$ (increased spatial sigma)
    \item Scenario 4: $L=100$ km, $\sigma_\text{sp} = 2$, $\sigma_\text{ind} = 1.5$ (increased noise)
\end{enumerate}
These four scenarios allow us to investigate the influence of different spatial error structures on the accuracy of our estimation and inference procedures. For each simulation scenario, we generate 100 simulated datasets (using 100 different seeds for random number generation), each having the same true parameter values and (real) exposure and covariates but different synthetic outcome variables.

Note that both Scenario 2 (increasing $L$) and Scenario 3 (increasing $\sigma_\text{sp}$) generate greater similarity in the value of outcomes that are spatially proximate, through different mechanisms, relative to Scenario 1. Whereas, Scenario 4 simply increases the random noise in the outcomes.

\subsection{Estimating the Effect Curve and Confidence Interval}

When implementing our ADT estimation procedure on each simulated dataset, we correctly specify the outcome models' mean function (just the variables, not the coefficients) and estimate the propensity score models $\pi_B$ and $\pi_D$ using Ranger (the same algorithm as in our main analysis). Cross-fitting is again used for $\mu_1$ and $\pi_D$.

When applying the Spatial Weighted Bootstrap, we fit both Exponential and Power covariance functions (with corresponding correlation functions $Corr(m; L) = exp(-m/L)$ and $Corr(m; a, z) = 1 - zm^a$, respectively) to the empirical variogram of the residuals from the LLKR stage and use whichever model has the smaller RMSE to generate the spatial weights. As noted in Section \ref{sec:residual-corr}, the spatial structure of the LLKR residuals is different than the spatial structure of the outcome (here, $\tilde{Y}$), so traditional notions of correctly vs incorrectly specifying the spatial correlation model do not directly apply here.

\subsection{Evaluating the Performance of the Spatial and Non-spatial Weighted Bootstraps}

We compare our ADT curve estimates from each simulated dataset to the true ADT curve by computing the pointwise bias and MAE, which are then summarized by averaging across all 100 simulation replicates within each scenario. We then calculate the pointwise coverage and width of the 95\% confidence intervals computed using the Spatial Weighted Bootstrap, and compare the results with  those from a Non-spatial Weighted Bootstrap.

\subsection{Simulation Results and Discussion}

\begin{figure}[!h] 
    \centering
    \includegraphics[width=0.8\textwidth]{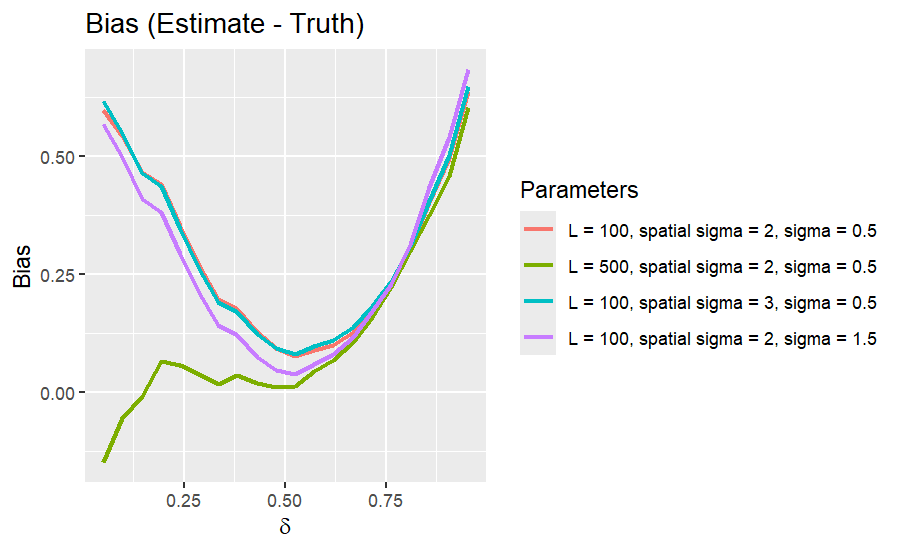} 
    \caption{Bias of the estimated ADT compared to the true ADT in our simulations.}
    \label{fig:Bias}  
\end{figure}

\begin{figure}[!h] 
    \centering
    \includegraphics[width=0.8\textwidth]{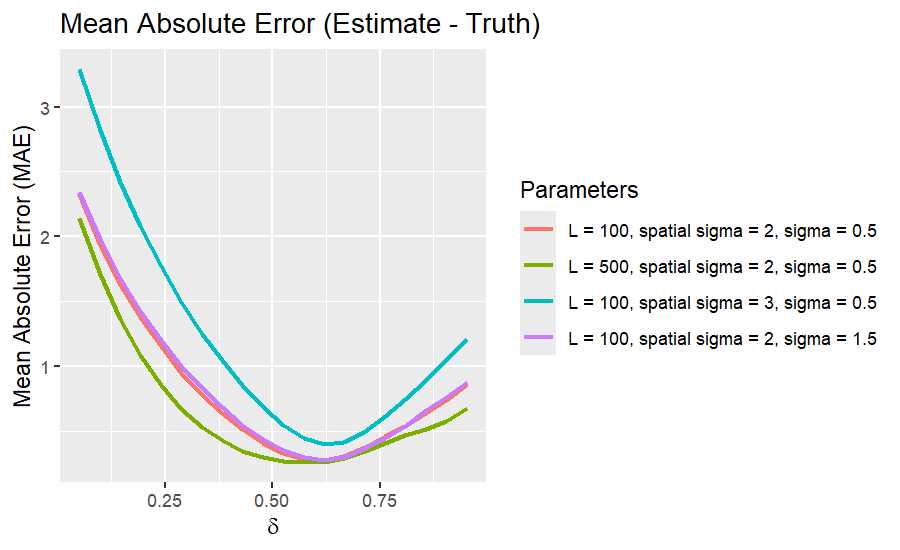} 
    \caption{Mean Absolute Error (MAE) of the estimated ADT compared to the true ADT in our simulations.}
    \label{fig:MAE}  
\end{figure}

\begin{figure}[!h] 
    \centering
    \includegraphics[width=0.9\textwidth]{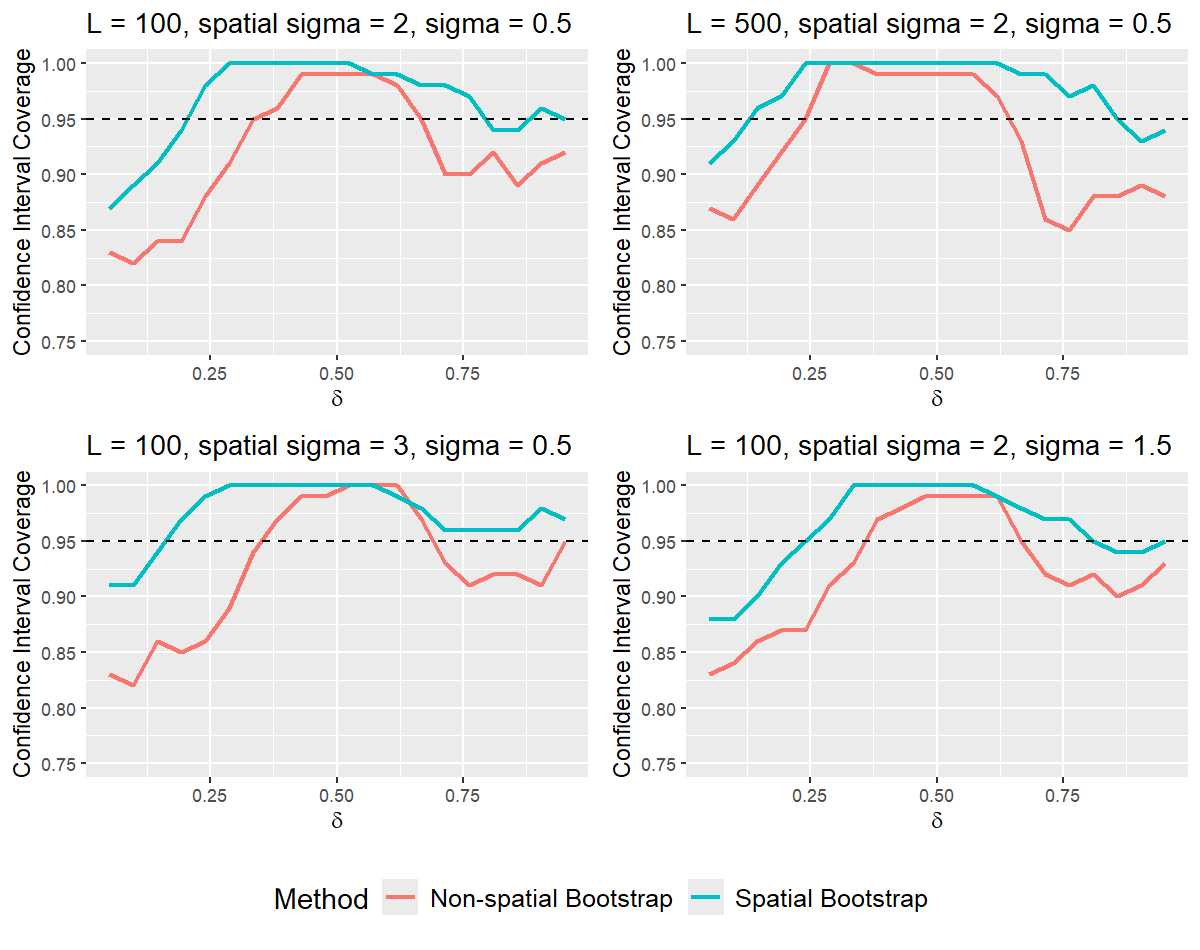} 
    \caption{Coverage of the 95\% confidence intervals in our simulations. The dashed line indicates the desired 95\% coverage.}
    \label{fig:coverage}  
\end{figure}

\begin{figure}[!h] 
    \centering
    \includegraphics[width=0.9\textwidth]{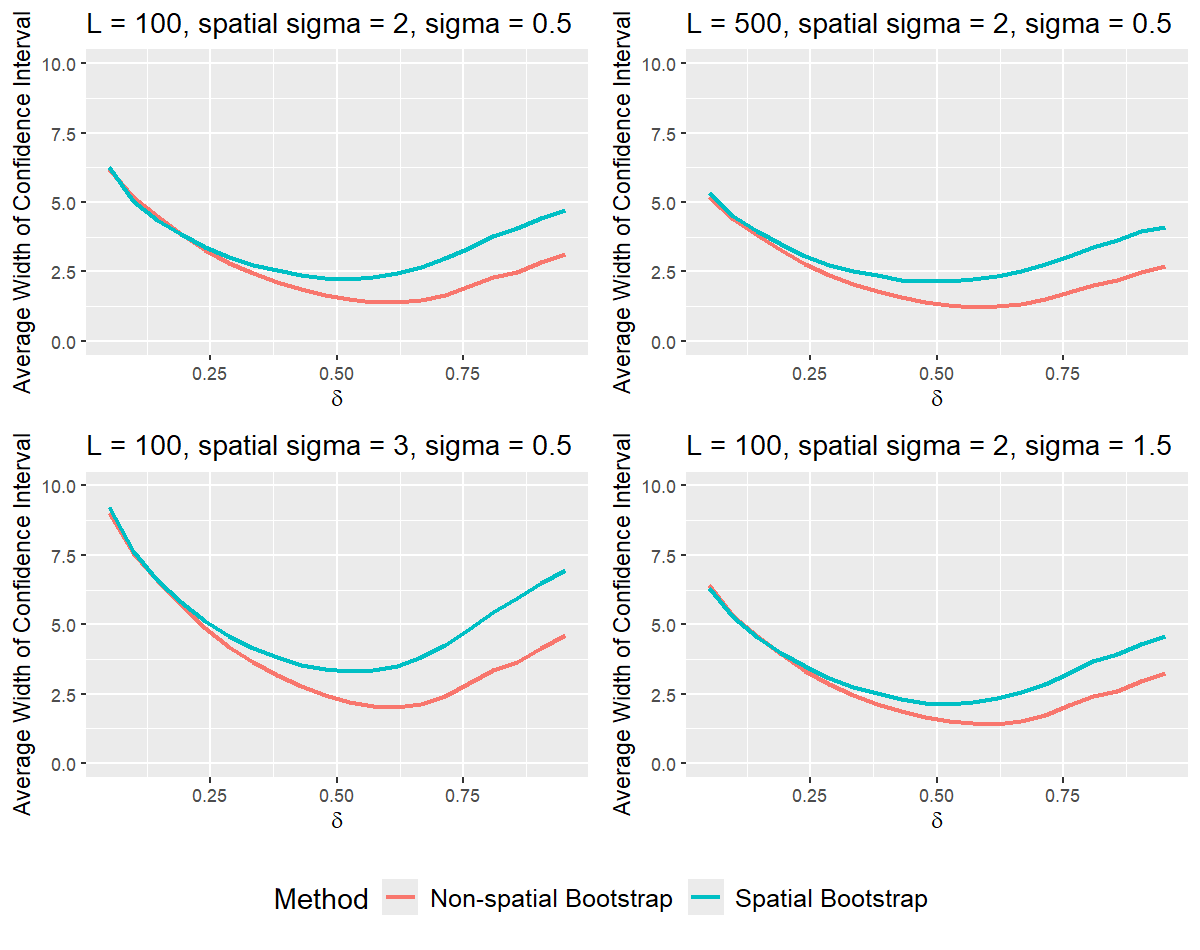} 
    \caption{Width of the confidence intervals in our simulations.}
    \label{fig:width}  
\end{figure}

The bias and MAE results are shown in Figures \ref{fig:Bias} and \ref{fig:MAE}, respectively. The confidence interval coverage and width results are shown in Figures \ref{fig:coverage} and \ref{fig:width}, respectively. 

First, we observe that the magnitude of the bias ranges between -0.15 and 0.68, while the true ADT ranges between 0.1 and 2.6, with an average value of 1.9. For all scenarios except the one with increased spatial range, the bias is largest on the tails. For the increased spatial range scenario, the bias becomes negative at low values of port proximity.

From the MAE plot, we see that the estimator is less stable on the tails (resulting in larger errors), which is to be expected for methods like LLKR. We observe that while changing the magnitude of noise does not impact the MAE, increasing $\sigma_{sp}$ results in higher MAE across all values of port proximity. To a lesser extent, increasing the spatial range results in a smaller MAE. 

\begin{figure}[!h] 
    \centering
    \includegraphics[width=0.8\textwidth]{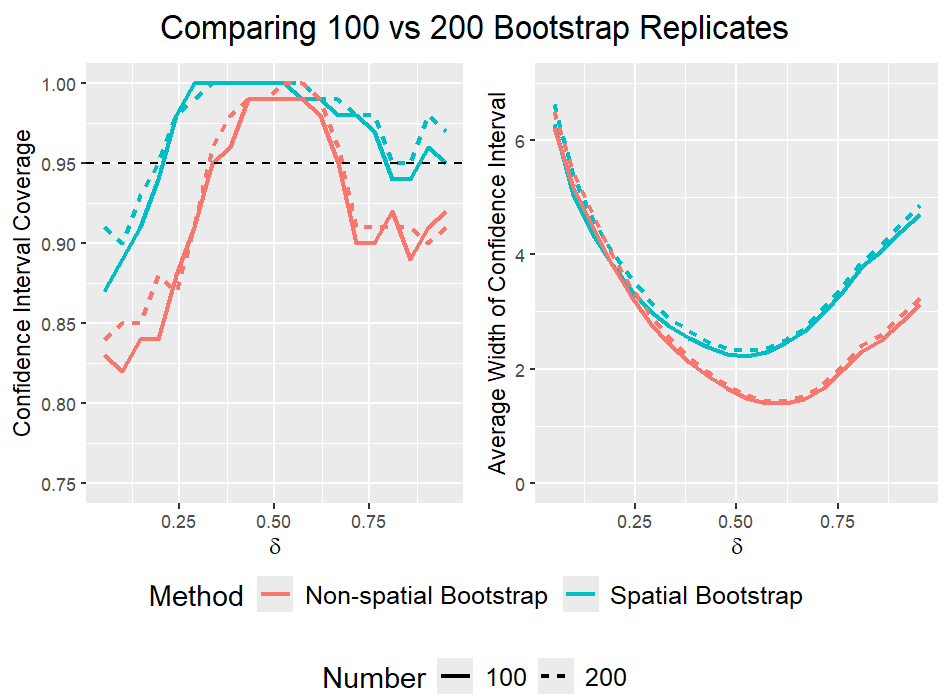} 
    \caption{Coverage and width of the confidence intervals when using 100 vs 200 bootstrap replicates.}
    \label{fig:num_boots}  
\end{figure}

In Figure \ref{fig:coverage}, we see that the Spatial Weighted Bootstrap has consistently higher coverage than the Non-Spatial Weighted Bootstrap, in part due to having wider confidence intervals as shown in Figure \ref{fig:width}. Mirroring the results of the MAE and Bias plots, both approaches have higher than 95\% coverage in the middle of the port proximity distribution, despite their confidence intervals narrowing. However, at low values of port proximity both methods have less-than-nominal coverage. At high values of port proximity, the Spatial Weighted Bootstrap outperforms the Non-Spatial approach, due to having wider confidence intervals. These results are stable across different numbers of bootstrap replicates (Figure \ref{fig:num_boots}).
Comparing across scenarios: we see that adding more spatially-structured noise increases the width of the confidence intervals. By contrast, the confidence intervals narrow slightly when the spatial range is increased. Similar to MAE, increasing the magnitude of random noise does not appear to change the interval widths. 

\begin{figure}[!h] 
    \centering
    \includegraphics[width=0.9\textwidth]{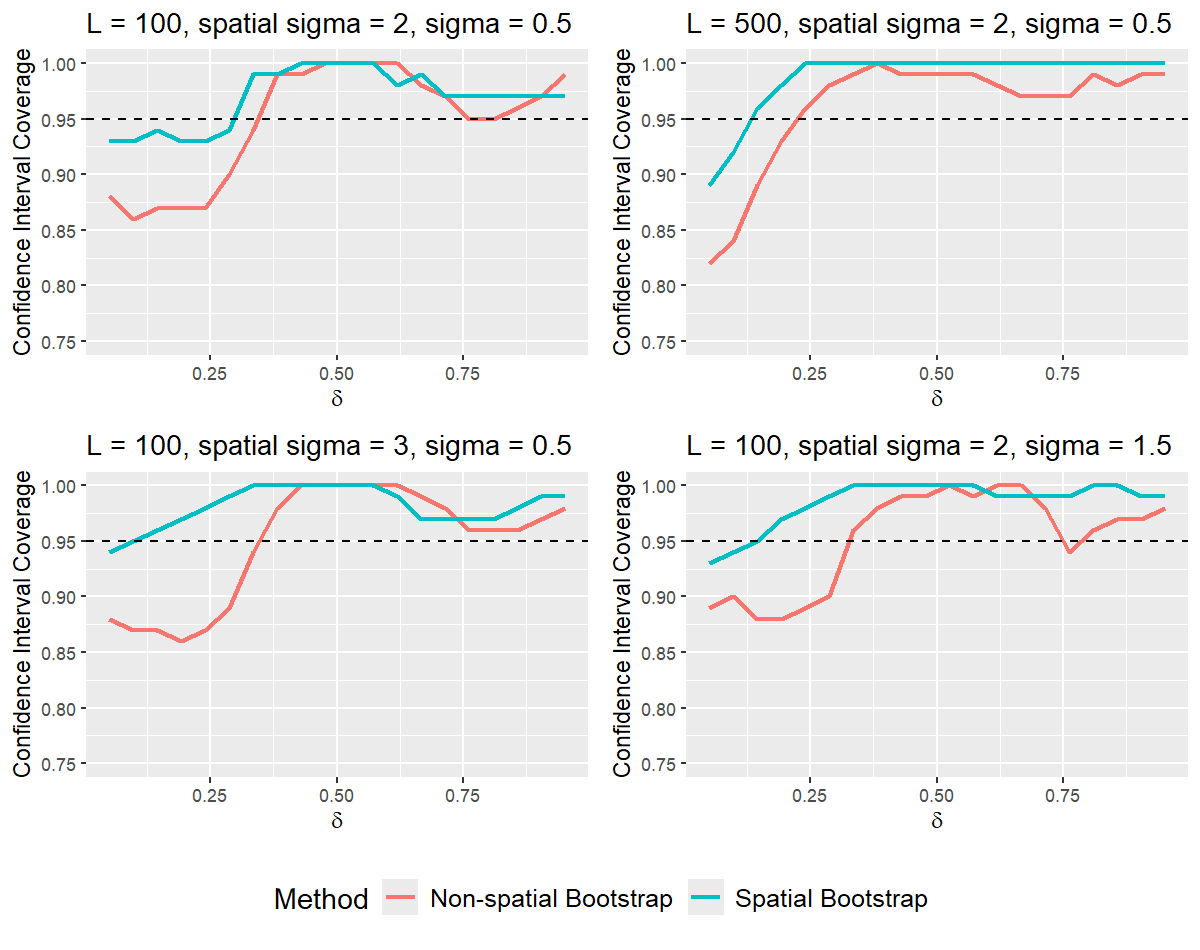} 
    \caption{Coverage of the 95\% confidence intervals in our simulations, \textit{after removing the average bias}. The dashed line indicates the desired 95\% coverage.}
    \label{fig:coverage-debiased}  
\end{figure}

To elucidate the extent to which under-coverage on the tails is driven by the finite sample bias of LLKR, we did as suggested by \citet{hettinger_MR} and evaluated the coverage after subtracting out the average bias. Figure \ref{fig:coverage-debiased} indicates that especially at higher values of port proximity, the LLKR bias is playing a substantial role in the coverage for both the Spatial and Non-Spatial Weighted Bootstrap approaches. At lower values of port proximity, the coverage gets pulled closer to 95\% for all scenarios except that with increased spatial range.

\end{appendix}

\end{document}